\documentclass[twocolumn,pre,showpacs]{revtex4}
\usepackage{graphicx}
\usepackage{amsmath}
\usepackage{amsfonts}
\usepackage{amssymb}
\usepackage{color}
\usepackage{float}
\usepackage{mathrsfs}

\begin{document}
\title{Integrable turbulence developing from strongly nonlinear partially coherent waves}

\author{D.\,S.~Agafontsev$^{1}$, S.~Randoux$^2$, P.~Suret$^2$}

\affiliation{\textit{$^1$ P.P. Shirshov Institute of Oceanology of RAS, 36 Nakhimovsky prosp., Moscow 117997, Russia.\\
$^2$ Laboratoire de Physique des Lasers, Atomes et Molecules, UMR-CNRS 8523, Universit\'e de Lille, Cit\'e Scientifique, 59655 Villeneuve d'Ascq Cedex, France.}}
\email{dmitrij@itp.ac.ru}

\begin{abstract}
We study numerically the integrable turbulence developing from strongly nonlinear partially coherent waves, in the framework of the focusing one-dimensional nonlinear Schr{\"o}dinger equation. 
We find that shortly after the beginning of motion the turbulence enters a state characterized by a very slow evolution of statistics (the quasi-stationary state -- QSS), and we concentrate on the detailed examination of the basic statistical functions in this state depending on the shape and the width of the initial spectrum. 
In particular, we show that the probability density function (PDF) of wavefield intensity is nearly independent of the initial spectrum and is very well approximated by a certain Bessel function representing an integral of the product of two exponential distributions. 
The PDF corresponds to the value of the second-order moment of intensity equal to $4$, indicating enhanced generation of rogue waves. 
All waves of large amplitude that we have studied are very well approximated -- both in space and in time -- by the rational breather solutions of either the first (the Peregrine breather), or the second orders. 
\end{abstract}

\maketitle


\section{Introduction}
\label{Sec:Intro}

The phenomenon of rogue waves -- extremely large waves that appear unpredictably from moderate wave background -- was intensively studied in the recent years, see e.g. the reviews~\cite{kharif2003physical,dysthe2008oceanic,onorato2013rogue}. 
One of the simplest nonlinear mathematical models suitable for the description of such waves is the one-dimensional nonlinear Schr{\"o}dinger equation (1D-NLSE) of the focusing type,
\begin{equation}\label{NLSE}
i\psi_t + \psi_{xx} + |\psi|^2 \psi=0,
\end{equation}
where $t$ is time, $x$ is spatial coordinate and $\psi$ is the wavefield envelope. 
Several exact solutions of this equation were suggested as candidates for rogue waves, including the Peregrine~\cite{peregrine1983water} and the Akhmediev~\cite{akhmediev1987exact,akhmediev2009rogue} breathers, the Kuznetsov-Ma solitons~\cite{kuznetsov1977solitons,ma1979perturbed} and the superregular breathers~\cite{gelash2014superregular}. 
Taking specific and carefully designed initial conditions, these solutions were reproduced in well-controlled experiments performed in different physical systems~\cite{kibler2010peregrine,chabchoub2011rogue,bailung2011observation,chabchoub2012super,chabchoub2012observation,kibler2012observation,kibler2015superregular}. 

With what frequency such solutions appear in nature, however, is a different question; for instance, typical oceanic waves represent weakly nonlinear objects having nearly Gaussian statistics, see e.g.~\cite{onorato2001freak,dysthe2008oceanic,onorato2013rogue}. 
This implies that the problem of rogue waves should be examined in the context of random initial conditions, and the key characteristic for the study should be the probability density function (PDF) of waves' amplitude, or, alternatively, the PDF $\mathcal{P}(I)$ of relative wave intensity $I=|\psi|^{2}/\langle|\psi|^{2}\rangle$, where $\langle|\psi|^{2}\rangle$ is the average intensity. 
Of particular interest is the comparison of the PDF for the nonlinear system with the exponential distribution
\begin{equation}\label{Rayleigh}
\mathcal{P}_{R}(I) = e^{-I},
\end{equation}
describing the PDF for a superposition of a multitude of uncorrelated linear waves with random phases, see e.g.~\cite{nazarenko2011wave}. 
In nonlinear systems, the phases of the Fourier modes composing the wavefield may correlate, that in turn may lead to enhanced appearance of large waves. 
Throughout the paper, we use the distribution~(\ref{Rayleigh}) as a benchmark allowing us to compare the frequency of large wave events for the examined system with that for a linear one. 

The 1D-NLSE is integrable in terms of the inverse scattering transform (IST)~\cite{zakharov1972exact,novikov1984theory}. 
Statistical analysis of integrable systems with random input is the concept of \textit{integrable turbulence}, first introduced by V.\,E. Zakharov in~\cite{zakharov2009turbulence}. 
Integrability implies conservation of infinite series of invariants (integrals of motion), and since these invariants are different for different types of initial conditions, the long-time wave statistics should strongly depend on the statistics of the input random process. 

In particular, for the scenario of noise-induced modulational instability (MI) of a plane wave (the condensate), the PDF of relative wave intensity coincides in the long time~\cite{agafontsev2015integrable,kraych2019statistical} with the exponential distribution~(\ref{Rayleigh}). 
This result was later generalized for the MI of cnoidal waves~\cite{agafontsev2016integrable}, with the demonstration that (depending on the parameters of the unstable cnoidal wave) either the PDF is sufficiently close to the exponential distribution, or the dynamics reduces to two-soliton collisions which occur with exponentially small rate. 
For both the condensate and the cnoidal wave initial conditions, the development of the MI leads to integrable turbulence, which approaches asymptotically to its \textit{stationary state} -- the state in which its statistical characteristics are independent of time. 

Meanwhile, for the partially coherent wave initial conditions having the exponential distribution of intensity~(\ref{Rayleigh}), the results were shown to be completely different~\cite{walczak2015optical}. 
The integrable turbulence in this case quickly reaches the state in which its statistical characteristics do not change visibly, and the PDF in this state exceeds the exponential distribution~(\ref{Rayleigh}) by orders of magnitude at large intensities (see also the earlier studies~\cite{onorato2000occurrence,onorato2004observation,onorato2006extreme} for the long-crested water waves of JONSWAP spectrum). 
Moreover, as was demonstrated in~\cite{suret2016single,tikan2018single,copie2019physics}, the excess over the exponential PDF is larger for the initial conditions of larger average intensity. 

In the present paper, with the help of extensive numerical simulations, we study in detail the special case of the latter scenario, namely, when the integrable turbulence develops from partially coherent wave source of high nonlinearity.
In terms of~\cite{suret2016single}, this case corresponds to large initial intensity, and, hence, appears to be one of the most promising candidate for the enhanced generation of rogue waves. 
As a measure of nonlinearity, we use the ratio 
\begin{equation}\label{alpha}
\alpha = \frac{|H_{nl}|}{H_{l}},
\end{equation}
of the potential energy $H_{nl}$ (related to the nonlinear term of the 1D-NLSE) to the kinetic one $H_{l}$ (related to dispersion), see Eqs~(\ref{energy-2})-(\ref{energy-3}) below. 
For practical applications, high nonlinearity can be achieved not only by increasing average intensity, but also by decreasing spectral width. 
In particular, a plane wave, being an example of wavefield with finite intensity and (in an ideal situation) infinite nonlinearity, is easily reproduceable in laboratory conditions. 
Real physical systems, however, have noise, and its effect may be significant. 
For this reason, we also study the influence of an additional wide-spectrum noise on the statistics. 

For partially coherent wave source of high nonlinearity (without additional wide-spectrum noise), we show that the developing integrable turbulence relatively quickly enters a \textit{quasi-stationary state} (QSS) -- the state in which some of its statistical characteristics (e.g., the wave-action spectrum and the PDF of intensity) change with time very slowly, while the other characteristics, most notably the autocorrelation of intensity, continue to change with time noticeably. 
The subsequent evolution towards the asymptotic \textit{stationary} state turns out to be very long, and we focus instead on the examination of the basic statistical functions in the beginning of the QSS. 
In particular, we demonstrate that, among these characteristics, the wave-action spectrum is symmetric, decays slightly slower than exponential at large wavenumbers and depends only on the level of nonlinearity of the initial conditions $\alpha_{0}=\alpha|_{t=0}$, but not on the shape of their spectrum -- even when the latter is rather generic and non-symmetric. 
The PDF exceeds the exponential distribution~(\ref{Rayleigh}) by orders of magnitude at large intensities, does not depend on the shape of the initial spectrum and slightly varies with the level of nonlinearity $\alpha_{0}$. 
The close-to-universal profile of the PDF is very well approximated by a certain Bessel function representing an integral of the product of two exponential distributions and corresponding to the value of the second-order moment of intensity $\kappa_{4} = \langle|\psi|^{4}\rangle/\langle|\psi|^{2}\rangle^{2} = 4$. 
The latter is indeed observed in the numerical experiments, indicating strong presence of rogue waves. 
We detect rogue waves exceeding the average amplitude by more than $14$ times, and all of the largest rogue waves that we have studied are very well approximated -- both in space and in time -- by the rational breather solutions of either the first (the Peregrine breather), or the second orders. 
The autocorrelation of intensity contains a universal bell-shaped central part with width of unity order, and also a non-universal part at intermediate distances between the universal central peak and large distances where the autocorrelation reaches unity. 
The non-universal part depends significantly on both the nonlinearity $\alpha_{0}$ and the shape of the initial spectrum. 
It also changes with time substantially, as well as the wave-action spectrum at large wavenumbers, the PDF at very large intensities and the moments of amplitude of higher order, revealing the hidden evolution in the QSS. 
Inclusion of an additional wide-spectrum noise to the initial conditions influences the transient regime preceding the QSS and, in the QSS, the wave-action spectrum and the kinetic energy. 
However, even significant noise levels practically do not change the moments, the PDF and the autocorrelation of intensity, opening perspectives for their experimental observation. 

The paper is organized as follows. 
In the next Section we give a brief overview for the problem of integrable turbulence developing from partially coherent wave source of high nonlinearity. 
In Section~\ref{Sec:NumMethods} we describe our numerical methods. 
In Section~\ref{Sec:Results1} we discuss the basic features of evolution from which we conclude the existence of the QSS, and in Section~\ref{Sec:Results2} we report our results for its statistical properties. 
In Section~\ref{Sec:Results3} we study the effect of inclusion of additional wide-spectrum noise to the initial conditions. 
The final Section contains conclusions. 
The paper has also Appendix~\ref{Sec:AppA}, where we describe construction of initial conditions with generic (non-symmetric) Fourier spectrum. 


\section{Partially coherent wave source of high nonlinearity}
\label{Sec:Formulation}

Without loss of generality, we study evolution governed by the focusing 1D-NLSE~(\ref{NLSE}) from the initial conditions of unit average intensity, 
\begin{equation}\label{NLSE-IC}
\psi|_{t=0}=\psi_{0}(x),\quad \overline{|\psi_{0}|^{2}}=\frac{1}{L}\int_{-L/2}^{L/2}|\psi_{0}|^{2}\,dx = 1.
\end{equation}
Here the overline denotes spatial averaging, and for the numerical study we consider the periodic problem $x\in[-L/2,L/2]$ with a very large period $L\gg 1$. 
The initial conditions are given by a discrete sum of Fourier components, 
\begin{equation}\label{NLSE-IC-F}
\psi_{0}(x) = \sum_{k}\psi_{0k}e^{ikx},\quad \psi_{0k} = \sqrt{N_{k}}e^{i\phi_{k}},
\end{equation}
where $k=2\pi m/L$ is the wavenumber, $m\in\mathbb{Z}$ is integer, $N_{k}\ge 0$ is a (given) smooth function of $k$ and $\phi_{k}\in[0,2\pi)$ are random phases. 
We use the so-called random phase (RP) model, see e.g.~\cite{nazarenko2011wave}, in which only the phases $\phi_{k}$ are considered to be random and uncorrelated, and average our results over the ensemble of phases' realizations. 

As an integrable equation, the 1D-NLSE conserves an infinite series of integrals of motion, and the first three of these invariants are wave action (in our notations equals to average intensity)
\begin{equation}\label{wave-action}
N = \overline{|\psi|^{2}} = \frac{1}{L}\int_{-L/2}^{L/2}|\psi|^{2}\,dx = \sum_{k}|\psi_{k}|^{2},
\end{equation}
momentum
\begin{equation}\label{momentum}
P = \frac{i}{2L}\int_{-L/2}^{L/2}(\psi_{x}^{*}\psi-\psi_{x}\psi^{*})\,dx = \sum_{k}k|\psi_{k}|^{2},
\end{equation}
and total energy 
\begin{eqnarray}
&& E = H_{l} + H_{nl}, \label{energy-1}\\
&& H_{l} = \overline{|\psi_{x}|^{2}} = \frac{1}{L}\int_{-L/2}^{L/2}|\psi_{x}|^{2}\,dx = \sum_{k}k^{2}|\psi_{k}|^{2}, \label{energy-2}\\
&& H_{nl} = -\frac{\overline{|\psi|^{4}}}{2} = -\frac{1}{2L}\int_{-L/2}^{L/2}|\psi|^{4}\,dx. \label{energy-3}
\end{eqnarray}
Here $H_{l}$ is the kinetic energy (linear contribution), $H_{nl}$ is the potential energy (nonlinear contribution), and $\psi_{k}$ is the Fourier-transformed wavefield, 
$$
\psi_{k}(t) = \frac{1}{L}\int_{-L/2}^{L/2}\psi(t,x)\,e^{-ikx}\,dx.
$$
Under the initial conditions of high nonlinearity we understand such initial conditions, for which the ratio of the potential energy to the kinetic one~(\ref{alpha}) is large, $\alpha_{0}=(|H_{nl}|/H_{l})|_{t=0}\gg 1$. 

According to the central limit theorem, the PDF of intensity for the initial conditions~(\ref{NLSE-IC})-(\ref{NLSE-IC-F}) is the exponential distribution, 
\begin{equation}\label{NLSE-IC-PDF}
\mathcal{P}_{R}(|\psi_{0}|^{2}) = \exp(-|\psi_{0}|^{2}).
\end{equation}
This allows us to find the ensemble-averaged potential energy,
\begin{eqnarray}
&& \langle H_{nl}\rangle|_{t=0} = -\frac{1}{2}\langle\overline{|\psi_{0}|^{4}}\rangle = \nonumber\\
&& = -\frac{1}{2}\int_{0}^{+\infty}|\psi_{0}|^{4}\,\mathcal{P}_{R}(|\psi_{0}|^{2})\,d|\psi_{0}|^{2}=-1, \label{NLSE-IC-Hnl}
\end{eqnarray}
and the second-order moment of intensity (which we call below as the fourth-order moment of amplitude)
\begin{equation}\label{k4}
\kappa_{4}|_{t=0} = \frac{\langle\overline{|\psi_{0}|^{4}}\rangle}{\langle\overline{|\psi_{0}|^{2}}\rangle^{2}}=2,
\end{equation}
where $\langle ...\rangle$ means averaging over the ensemble of random phases. 
Hence, for high nonlinearity, the initial conditions must have small kinetic energy, $H_{l}\ll 1$. 
The quantity 
\begin{equation}\label{NLSE-IC-Dk}
\delta k^{2} = \frac{\sum_{k}k^{2}|\psi_{k}|^{2}}{\sum_{k}|\psi_{k}|^{2}} = \frac{H_{l}}{N},
\end{equation}
has the meaning of square spectral width, and for the initial conditions of unit average intensity, $N=1$, equals to the kinetic energy, $\delta k^{2}=H_{l}$, see Eqs.~(\ref{NLSE-IC}),~(\ref{wave-action}),~(\ref{energy-2}). 
Hence, high nonlinearity means small spectral width. 

To summarize, we examine evolution governed by the 1D-NLSE~(\ref{NLSE}) starting from random superpositions of uncorrelated linear waves~(\ref{NLSE-IC})-(\ref{NLSE-IC-F}) of high nonlinearity, $\alpha_{0}=|H_{nl}|/H_{l}\gg 1$. 
Such initial conditions are characterized by the exponential distribution of intensity~(\ref{NLSE-IC-PDF}), have small spectral width and kinetic energy, $\delta k^{2} = H_{l} = 1/\alpha_{0}\ll 1$, the potential energy of minus unity, $\langle H_{nl}\rangle = -1$, the total energy close to minus unity, $\langle E\rangle = -1+1/\alpha_{0}\approx -1$, and the fourth-order moment of amplitude that equals to two, $\kappa_{4} = 2$. 
In the physical space, each realization represents a collection of humps of characteristic width $\delta x\simeq 1/\delta k=\sqrt{\alpha_{0}}\gg 1$. 
The linear time-scale for the problem is large, $t_{l}=\delta x^{2}\simeq \alpha_{0}\gg 1$, and the nonlinear time-scale equals to unity, $t_{nl}=1$, see e.g.~\cite{agrawal2001nonlinear}. 

The described initial conditions resemble the condensate in the sense of small spectral width. 
However, importantly, the condensate cannot be obtained from them in the limit of large nonlinearity, since it has different total energy $E=-1/2$ and delta-distribution of intensity. 

Note that the problem of evolution from initial conditions of arbitrary average intensity, $\overline{|\Psi_{0}|^{2}}=N_{0}$, within the framework of the focusing 1D-NLSE
$$
i\Psi_{T} + \beta\,\Psi_{XX} + \gamma\,|\Psi|^2 \Psi = 0,
$$
with arbitrary dispersion $\beta>0$ and nonlinearity $\gamma>0$ coefficients, renormalizes to Eqs.~(\ref{NLSE})-(\ref{NLSE-IC}) with the help of the scaling transformations $\Psi=\sqrt{N_{0}}\,\psi$, $T=t/\gamma N_{0}$ and $X=x\sqrt{\beta/\gamma N_{0}}$. 
With these transformations, the wavenumber renormalizes as $K=k\sqrt{\gamma N_{0}/\beta}$, so that in the dimensionless variables the level of the initial nonlinearity $\alpha_{0}$ depends on the (original) spectral width $\delta K$ and the average intensity $N_{0}$ as $\alpha_{0}=\delta k^{-2}\propto N_{0}/\delta K^{2}$. 
This means that the nonlinearity can be enhanced by both increasing the average intensity $N_{0}$ and decreasing the spectral width $\delta K$. 
In particular, the experiments in~\cite{suret2016single} with larger initial intensities (optical power), which demonstrated the larger probability of rogue wave events, have in our formulation the smaller (renormalized) spectral width $\delta k$ and the larger level of nonlinearity $\alpha_{0}$.

Also note, that in our formulation the fourth-order moment equals to the double absolute value of the potential energy, $\kappa_{4} = 2|\langle H_{nl}\rangle|$, see Eqs.~(\ref{NLSE-IC}),~(\ref{energy-3}),~(\ref{k4}). 
When we start from initial conditions of small spectral width, $\delta k\ll 1$, we expect the latter to increase to some value of unity order during the evolution in time. 
This means increase in the kinetic energy $H_{l}$ to some value of unity order, see Eq.~(\ref{NLSE-IC-Dk}), and, due to the conservation of the total energy, the same increase in the absolute value of the potential energy, as well. 
The latter equals to unity at the initial time, $|\langle H_{nl}\rangle|=1$, that corresponds to the fourth-order moment $\kappa_{4} = 2$. 
Hence, with the increase of the spectral width during the evolution, we also expect the fourth-order moment to significantly increase from the value of two, that would correspond to strongly non-exponential PDF and enhanced probability of rogue wave events. 
These ideas exploiting the connection between the spectral width, the fourth-order moment and the probability of rogue wave appearance were first suggested in~\cite{onorato2016origin} and, as demonstrated in there, can be generalized for a class of NLS-type equations in multiple spatial dimensions. 

In the present paper, among the statistical characteristics of the system, we examine the ensemble-average kinetic $\langle H_{l}(t)\rangle$ and potential $\langle H_{nl}(t)\rangle$ energies, the moments of amplitude $M^{(p)}(t)=\langle\overline{|\psi|^{p}}\rangle/\langle\overline{|\psi|^{2}}\rangle^{p/2}\equiv \langle\overline{|\psi|^{p}}\rangle$ and, in particular, the fourth-order moment $\kappa_{4}=M^{(4)}$, the PDF $\mathcal{P}(I,t)$ of relative wave intensity $I=|\psi|^{2}/\langle\overline{|\psi|^{2}}\rangle\equiv |\psi|^{2}$, the wave-action spectrum, 
\begin{equation}\label{wave-action-spectrum}
S_{k}(t) = \langle|\psi_{k}|^{2}\rangle/\Delta k,
\end{equation}
where $\Delta k = 2\pi/L$ is the distance between neighbor harmonics, and the autocorrelation of intensity, 
\begin{equation}\label{g2}
g_{2}(x,t) = \frac{\langle\overline{|\psi(y+x,t)|^{2}|\psi(y,t)|^{2}}\rangle}{\langle\overline{|\psi(y,t)|^{2}}\rangle^{2}}.
\end{equation}
In the latter relation, the overline means averaging over the spatial coordinate $y$. 
Note that at $x=0$ the autocorrelation equals to the fourth-order moment, $g_{2}(0,t)=\kappa_{4}(t)$, and at $x\to\infty$ it must approach to unity, $g_{2}(x,t)\to 1$. 
For the wave-action spectrum and the PDF, we use normalization conditions $\int S_{k}\,dk = 1$ and $\int \mathcal{P}(I)\,dI = 1$, respectively. 
In the next Sections we will also use that the exponential PDF~(\ref{Rayleigh}) corresponds to the following values of the moments, 
\begin{equation}\label{moments-Rayleigh}
M^{(p)}_{R} = \int_{0}^{+\infty}|\psi|^{p}\,\mathcal{P}_{R}(|\psi|^{2})\,d|\psi|^{2} = \Gamma_{1+p/2},
\end{equation}
where $\Gamma$ is Gamma-function. 


\section{Numerical methods}
\label{Sec:NumMethods}

We integrate Eq.~(\ref{NLSE}) numerically in the box $x\in[-L/2, L/2]$, $L\ge 512\pi$, with periodic boundaries. 
We use the pseudospectral Runge-Kutta fourth-order method in adaptive grid, with the grid step $\Delta x$ set from the analysis of the Fourier spectrum of the solution, see~\cite{agafontsev2015integrable} for detail. 
The time step $\Delta t$ changes with $\Delta x$ as $\Delta t = h\,\Delta x^{2}$, $h\le 0.1$, in order to avoid numerical instabilities. 
We checked that the first ten integrals of motion of the 1D-NLSE are conserved by our numerical scheme up to the relative errors from $10^{-10}$ (the first three invariants) to $10^{-6}$ (the tenth invariant) orders. 

As initial conditions, we use partially coherent waves with super-Gaussian Fourier spectrum
\begin{eqnarray}
\psi_{0}(x) &=& \sum_{k} A_{k}^{(0)}\,e^{ikx+i\phi_{k}},\label{IC}\\
A_{k}^{(0)} &=& \bigg(\frac{C_{n}}{\theta L}\bigg)^{1/2}\,e^{-|k|^{n}/\theta^{n}},\label{IC-symmetric}
\end{eqnarray}
where $\theta$ plays the role of the spectral width~(\ref{NLSE-IC-Dk}), $\delta k\propto\theta$, $\phi_{k}$ are random phases for each $k$ and each realization of the initial conditions, $n\in\mathbb{N}$ is the exponent defining the shape of the initial spectrum, and $C_{n}=\pi\, 2^{1/n}/\Gamma_{1+1/n}$ is the normalization constant such that the average intensity is unity, $\overline{|\psi_{0}(x)|^{2}}=1$, see e.g. Eq.~(\ref{IC-Hd}) below and Eq.~(25) in~\cite{agafontsev2015integrable}. 
For each pair of parameters $n$ and $\theta$, we perform simulation for the ensemble of $1000$ random realizations of phases $\phi_{k}$ and then average the results over these realizations. 
The initial kinetic energy is proportional to $\theta^{2}$, 
\begin{eqnarray}
&& H_{l}|_{t=0}=\frac{1}{L}\int_{-L/2}^{+L/2}|\psi_{0x}|^{2}\,dx =\nonumber\\ 
&& = \frac{C_{n}}{\theta L}\sum_{k}k^{2}\,e^{-2k^{n}/\theta^{n}} \approx \frac{\Gamma_{1+3/n}}{\Gamma_{1+1/n}}\times\frac{\theta^{2}}{3\cdot 2^{2/n}}, \label{IC-Hd}
\end{eqnarray}
while the initial ensemble-averaged potential energy equals to minus unity, see Eq.~(\ref{NLSE-IC-Hnl}). 
Thus, the initial nonlinearity level (the potential-to-kinetic energy ratio) is inverse-proportional to $\theta^{2}$,
\begin{eqnarray}
\alpha_{0} = \frac{|\langle H_{nl}\rangle|}{\langle H_{l}\rangle} \approx \frac{\Gamma_{1+1/n}}{\Gamma_{1+3/n}}\times\frac{3\cdot 2^{2/n}}{\theta^{2}}.
\label{IC-ratio}
\end{eqnarray}

We perform simulations for several profiles of the initial spectrum~(\ref{IC-symmetric}), including exponential $n=1$, Gaussian $n=2$, and super-Gaussian $n=8$ and $n=32$, and also for several initial nonlinearity levels $\alpha_{0}$ from $1$ to $256$, with the parameter $\theta$ determined from Eq.~(\ref{IC-ratio}). 
We also study generic (non-symmetric) Fourier spectra $A_{k}^{(0)}$ in Eq.~(\ref{IC}), which we construct as described in Appendix~\ref{Sec:AppA}. 
The distance between the neighbor harmonics in our simulations depends only on the box size, $\Delta k = 2\pi/L$, and we keep the box sufficiently large, $L\ge 512\pi$, so that the region in the $k$-space $[-\theta, \theta]$, containing most of the initial wave action, see Eq.~(\ref{wave-action}), is resolved with
\begin{equation}\label{IC-resolution}
2\theta/\Delta k\gtrsim 100
\end{equation}
harmonics. 
We confirmed that simulations performed in twice larger boxes $L$ provide the same statistical results. 

\begin{figure*}[t]\centering
\includegraphics[width=5.15cm]{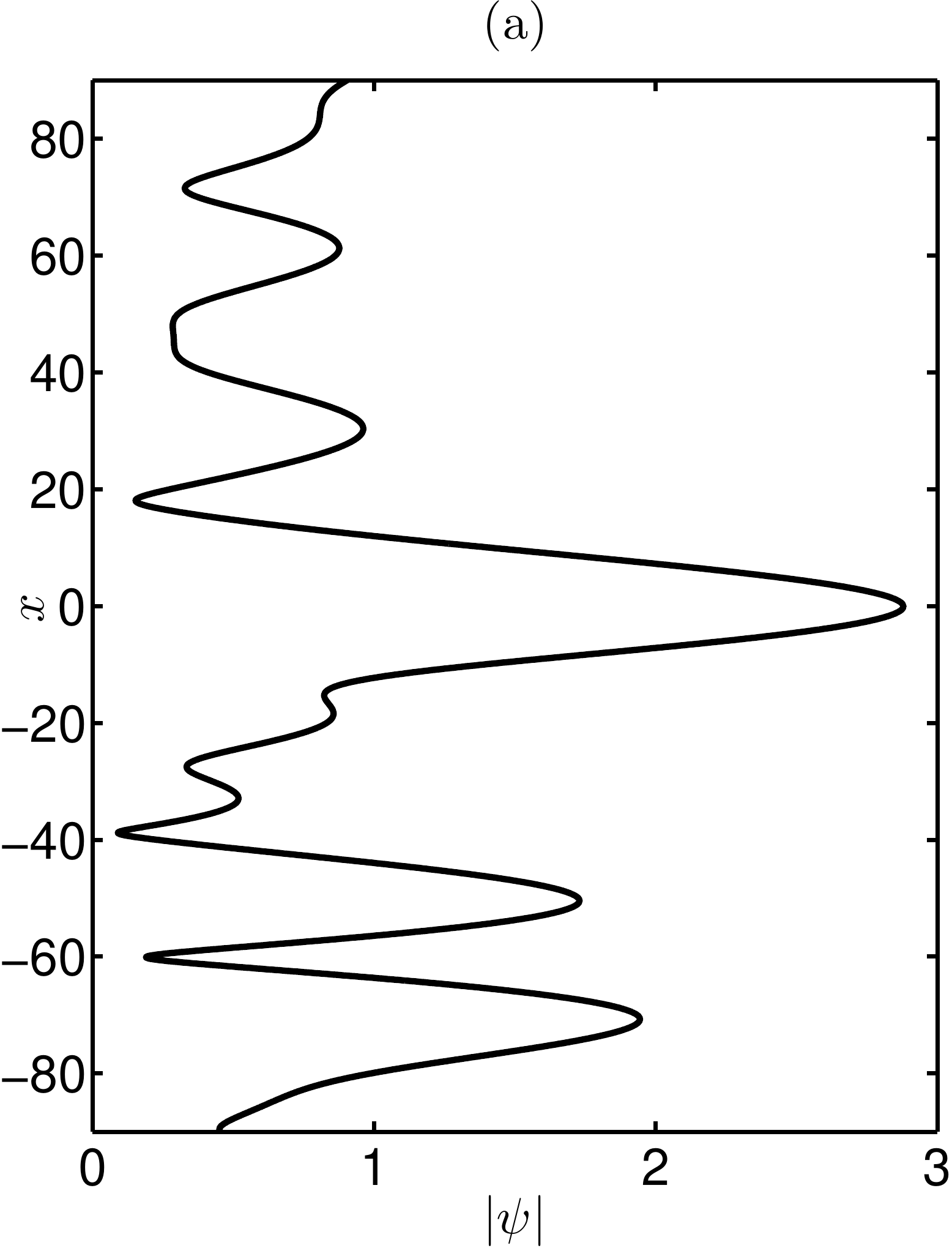}
\includegraphics[width=12.65cm]{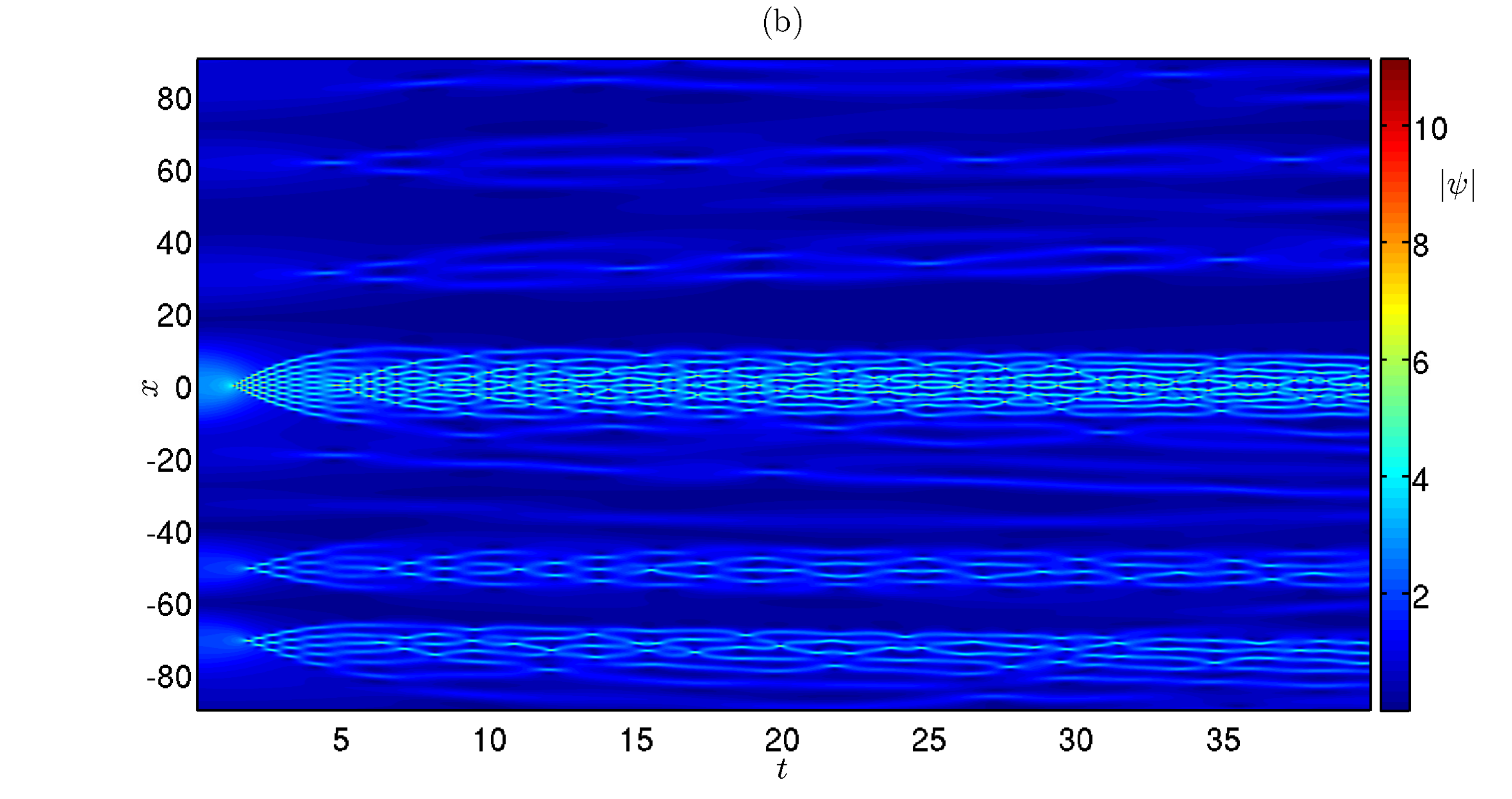}

\caption{\small {\it (Color on-line)} 
(a) Initial amplitude $|\psi|$ and (b) space-time evolution for one realization of partially coherent wave of high nonlinearity. 
The initial spectrum is super-Gaussian with the exponent $n=32$ and nonlinearity level $\alpha_{0}=64$. 
The maximum of the initial amplitude is shifted to $x=0$ for better visualization.
}
\label{fig:fig1}
\end{figure*}


\section{Time evolution and the quasi-stationary state}
\label{Sec:Results1}

We start this Section with the description of basic features of evolution from partially coherent wave of high nonlinearity that are known from previous publications, where, in addition to the 1D-NLSE in its original form, the semi-classical approximation was used as well. 
The latter starts with renormalization of the 1D-NLSE to the form, 
\begin{equation}
i\varepsilon\psi_{\tau} + \frac{\varepsilon^{2}}{2}\psi_{\zeta\zeta} + |\psi|^{2}\psi = 0, \label{NLSE-quasiclassical}
\end{equation}
where $\varepsilon = \sqrt{t_{nl}/t_{l}}\simeq 1/\sqrt{\alpha_{0}}$ is a small parameter, $t_{l}\gg t_{nl}$ are linear and nonlinear time scales, and $\zeta=\varepsilon\,x/\sqrt{2}$ and $\tau = \varepsilon\,t$ are renormalized space and time. 
The application of the Madelung transform separates the complex wavefield $\psi$ by the amplitude and the phase, and ultimately results in equations of motion for one-dimensional fluid~\cite{randoux2017optical}. 
The new equations contain terms proportional to $\varepsilon^{2}$, that for the initial conditions in the form of a single hump with width and height of unity order $\delta\zeta\sim 1$ and $|\psi|\sim 1$ can be neglected at the early stage of the evolution. 
However, without these terms at later stages, there exists a certain (critical) time of unity order $\tau\sim 1$ when the gradient of the solution explodes (the \textit{gradient catastrophe}). 
As was found in~\cite{bertola2013universality}, the full equation~(\ref{NLSE-quasiclassical}) at the critical time is regularized by the emergence of a local coherent structure, which tends asymptotically to the Peregrine breather solution~\cite{peregrine1983water} as $\varepsilon\to 0$.
The subsequent optical fiber experiments~\cite{tikan2017universality} have demonstrated that the described scenario with the coherent structure fitted locally by the Peregrine solution can be observed for the parameter $\varepsilon$ as large as $0.45$. 

As we have mentioned in Section~\ref{Sec:Formulation}, a partially coherent wave of high nonlinearity represents a collection of humps of characteristic width $\delta x\simeq \sqrt{\alpha_{0}}$. 
In renormalized variables $(\zeta,\tau)$, these humps have width and height of unity order $\delta\zeta\sim 1$ and $|\psi|\sim 1$. 
It is naturally to assume that, at the early stage of the evolution, the humps may be considered as independent objects. 
Then, each hump should undergo its own gradient catastrophe regularized by the emergence of the Peregrine-like coherent structure at its own critical time. 
A detailed numerical study with the demonstration of this scenario has been performed in~\cite{tikan2019effect}, and it has been shown that the time when the fourth-order moment $\kappa_{4}$ takes maximum value corresponds to the maximum presence of the Peregrine-like coherent structures appearing on top of different humps. 
The time of maximum fourth-order moment turns out to be also of unity order in the renormalized variables $\tau_{m}\sim 1$, so that in our variables $(x,t)$ it should scale with the nonlinearity level $\alpha_{0}$ as $t_{m}\sim\sqrt{\alpha_{0}}$. 

\begin{figure*}[t]\centering
\includegraphics[width=11.9cm]{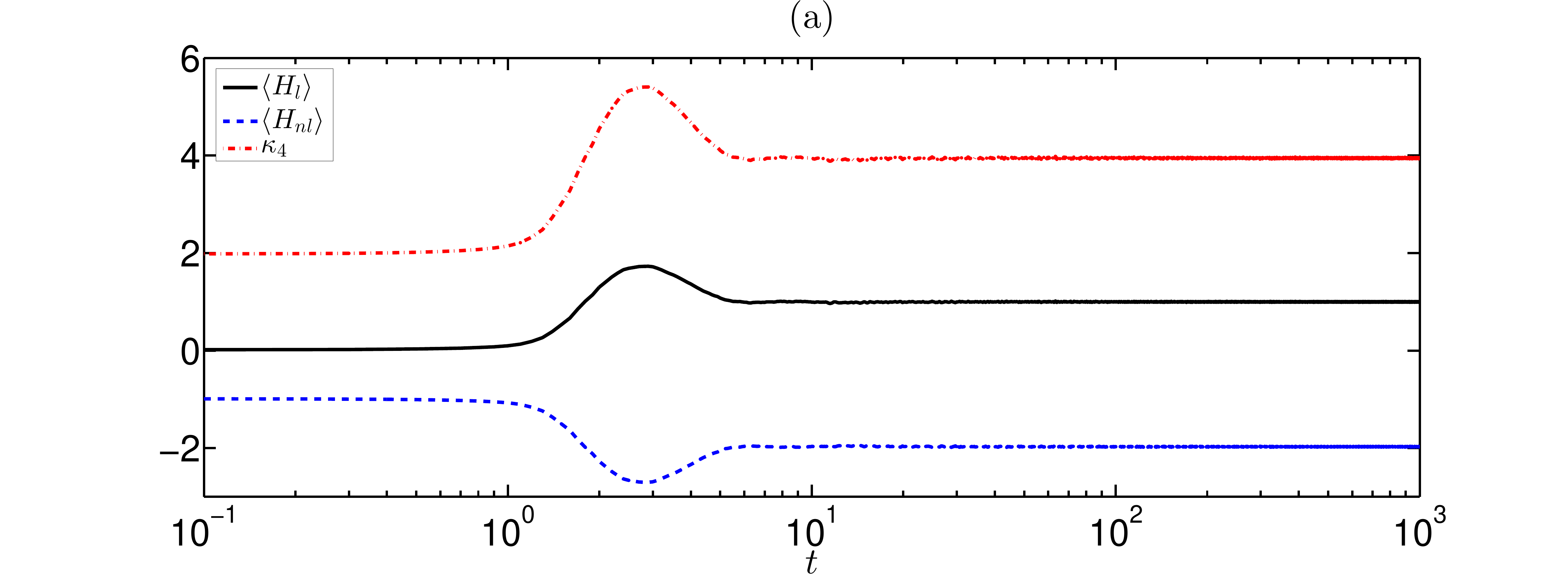}
\includegraphics[width=5.9cm]{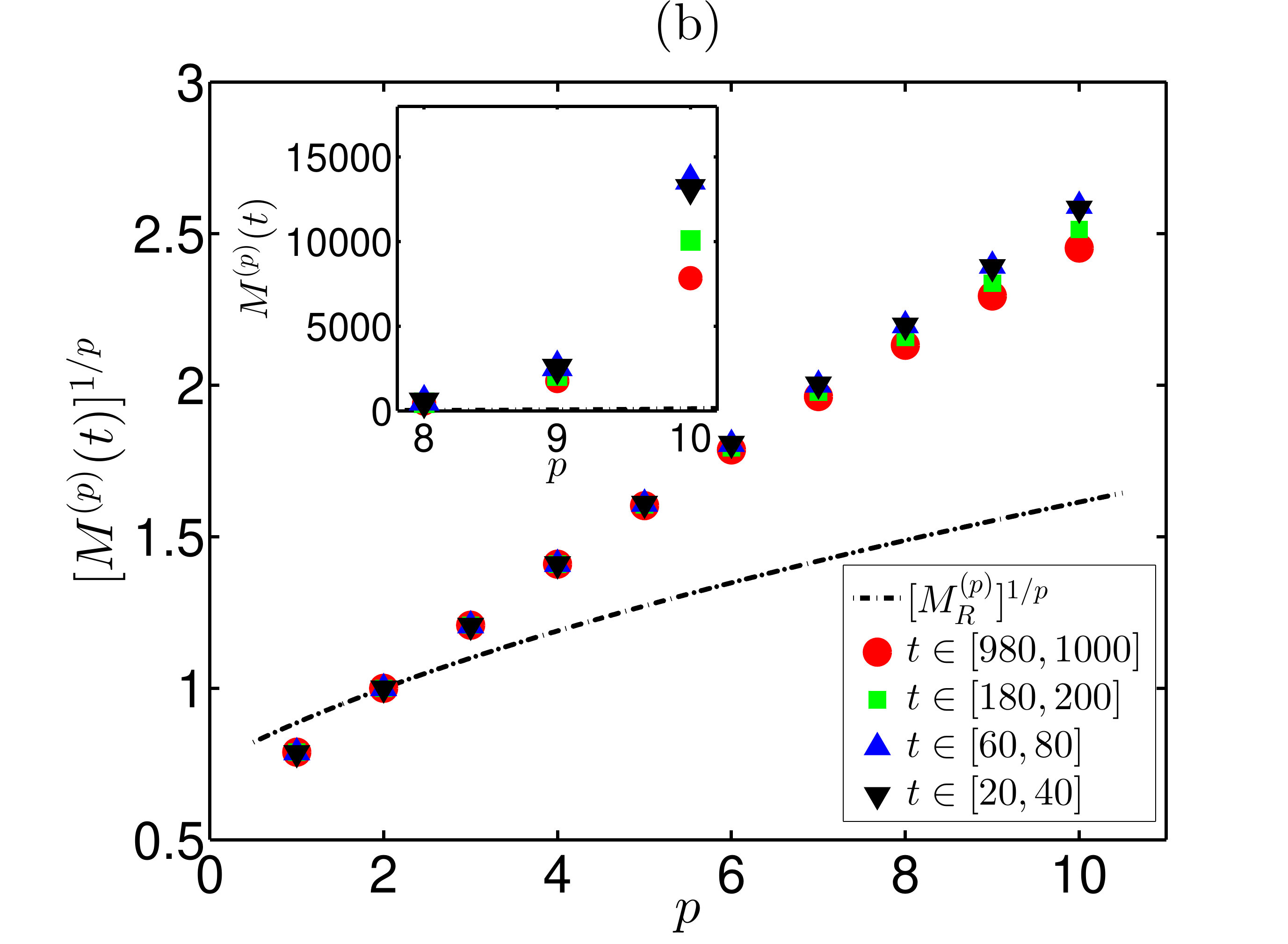}\\
\includegraphics[width=5.9cm]{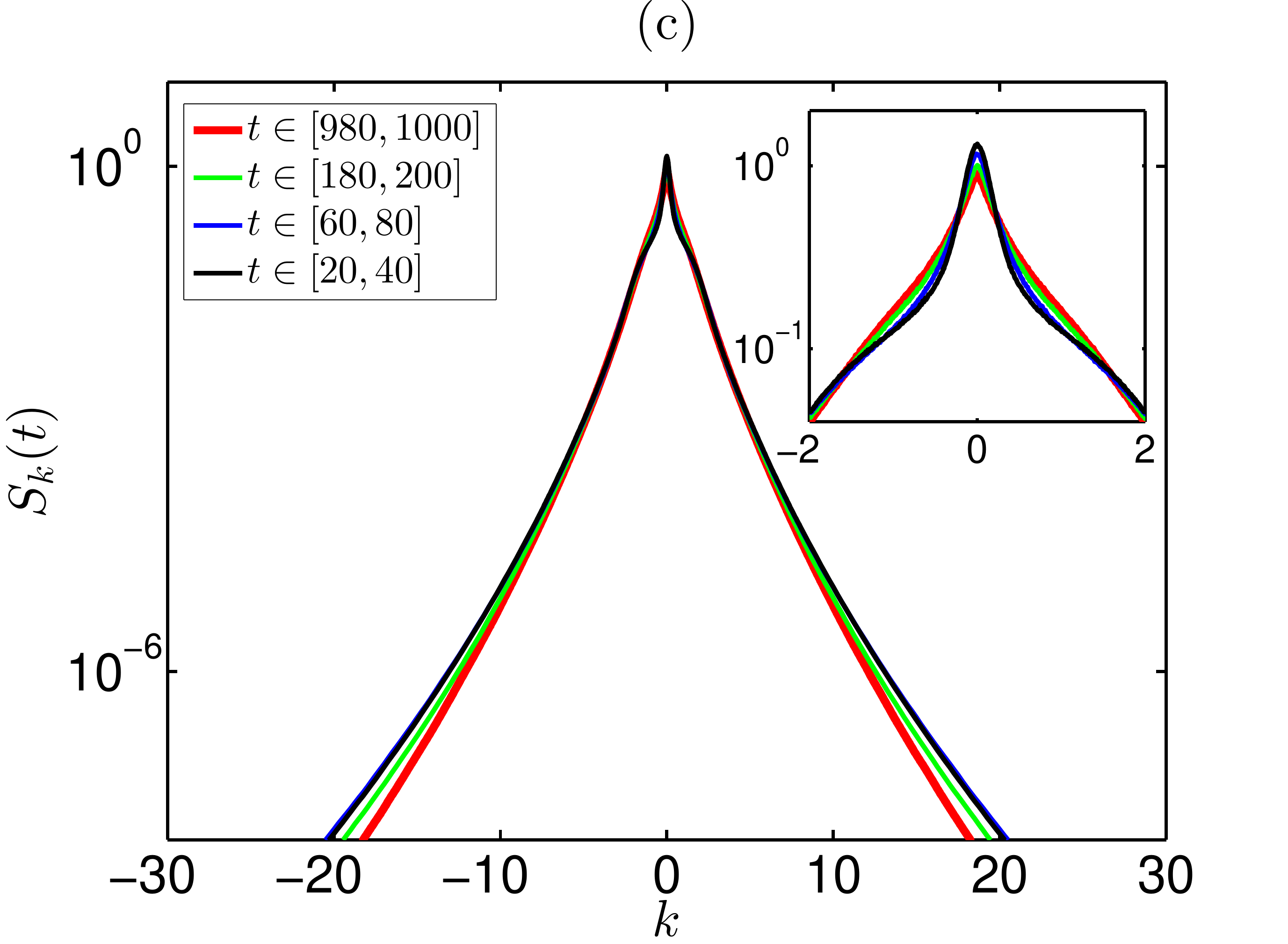}
\includegraphics[width=5.9cm]{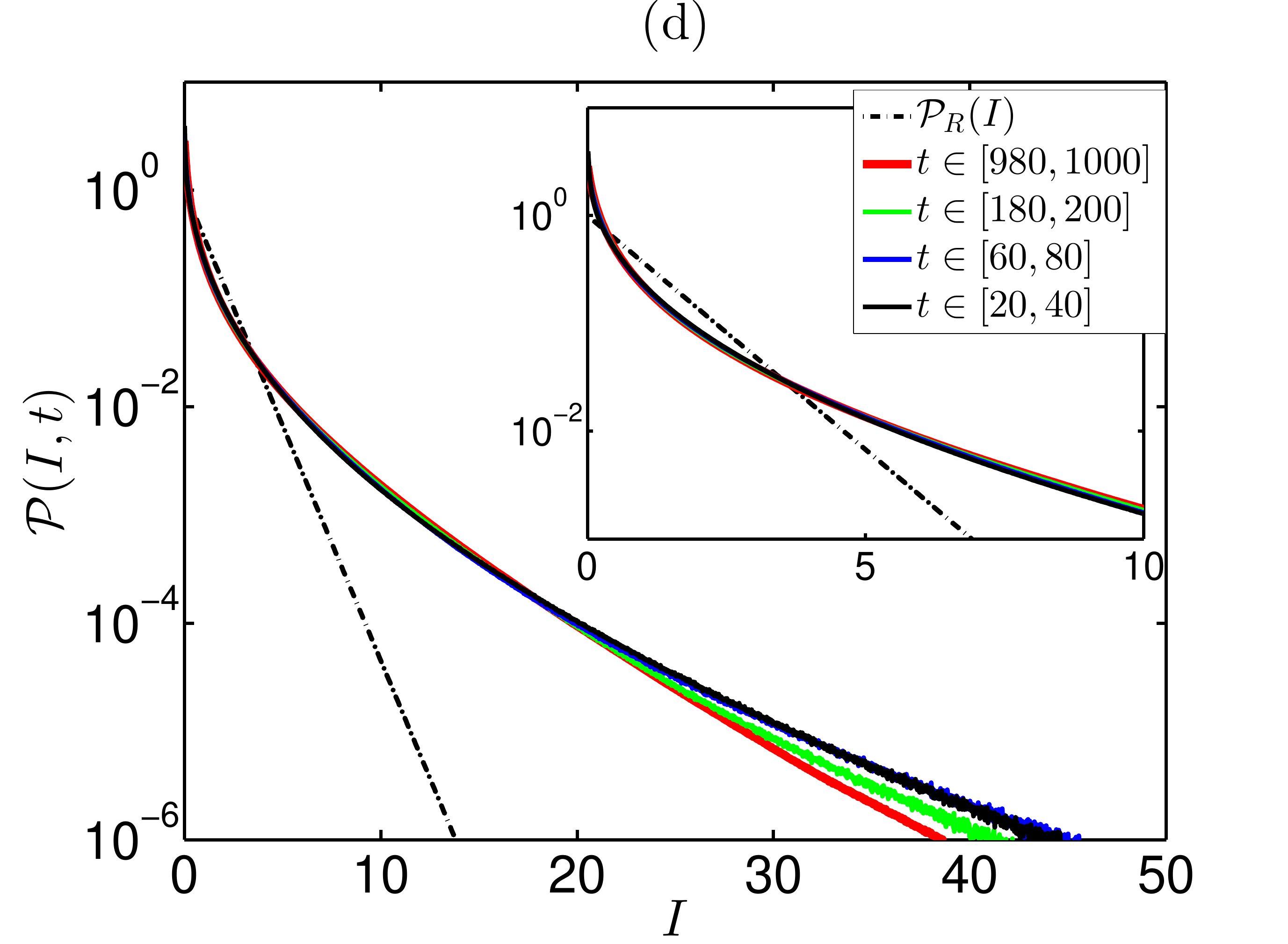}
\includegraphics[width=5.9cm]{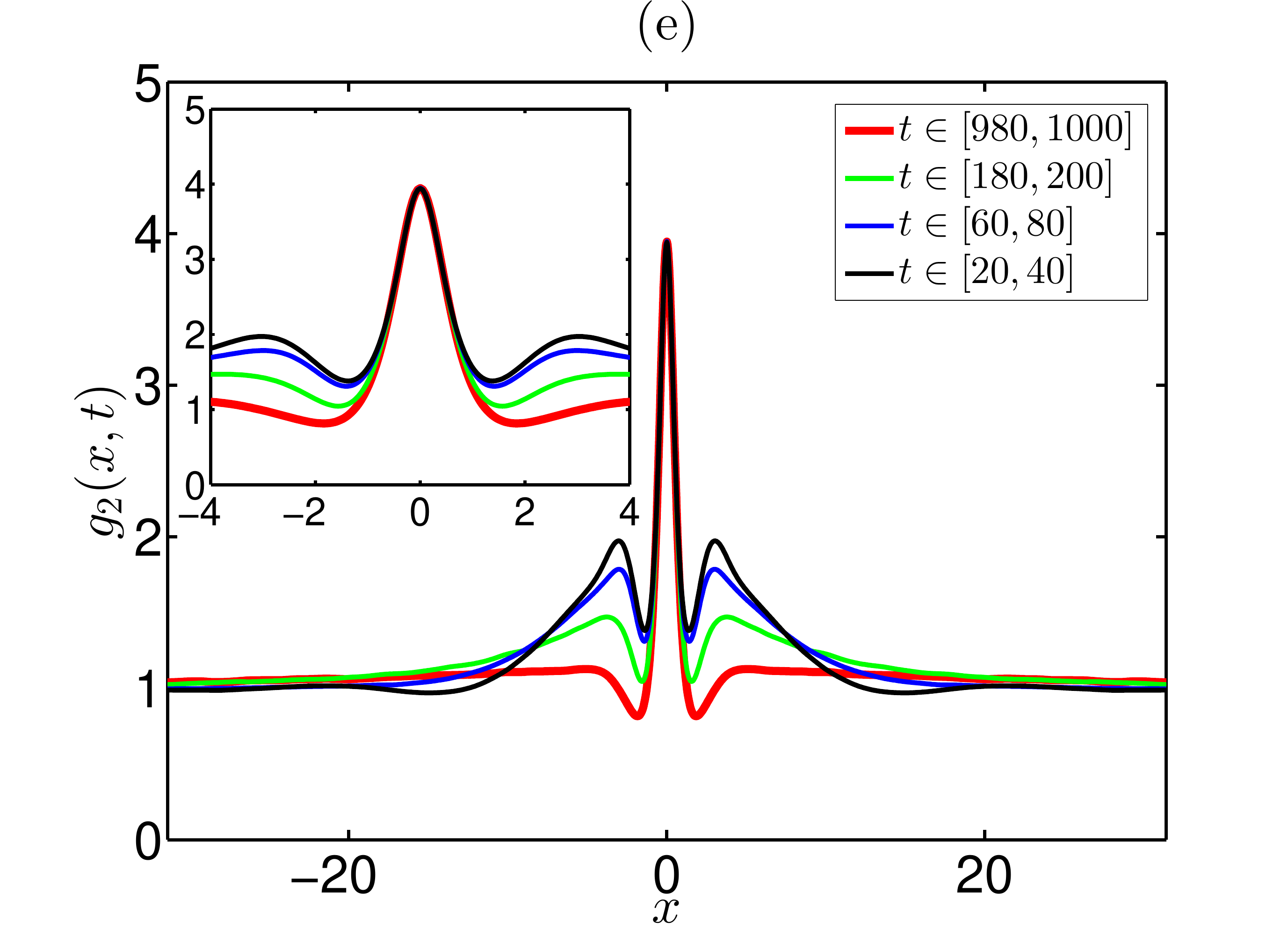}

\caption{\small {\it (Color on-line)} 
(a) Evolution of the ensemble-averaged kinetic energy $\langle H_{l}\rangle$, potential energy $\langle H_{nl}\rangle$ and the fourth-order moment $\kappa_{4}$; note the logarithmic horizontal scale. 
(b-e) Averaged over ensemble and different time intervals statistical functions: (b) the moments $[M^{(p)}]^{1/p}$, (c) the wave-action spectrum $S_{k}$, (d) the PDF $\mathcal{P}(I)$ and (e) the autocorrelation of intensity $g_{2}(x)$. 
The initial spectrum is super-Gaussian with the exponent $n=32$ and nonlinearity level $\alpha_{0}=64$. 
The inset in the panel (b) shows the higher-order moments $M^{(p)}$ (without the $1/p$ power), while the insets in the panels (c-e) -- the same functions as in the main figures with smaller scales. 
In the panels (d) and (b), the black dash-dot lines indicate the exponential PDF~(\ref{Rayleigh}) and the corresponding moments~(\ref{moments-Rayleigh}), respectively.
}
\label{fig:fig2}
\end{figure*}

\begin{figure*}[t]\centering
\includegraphics[width=8.9cm]{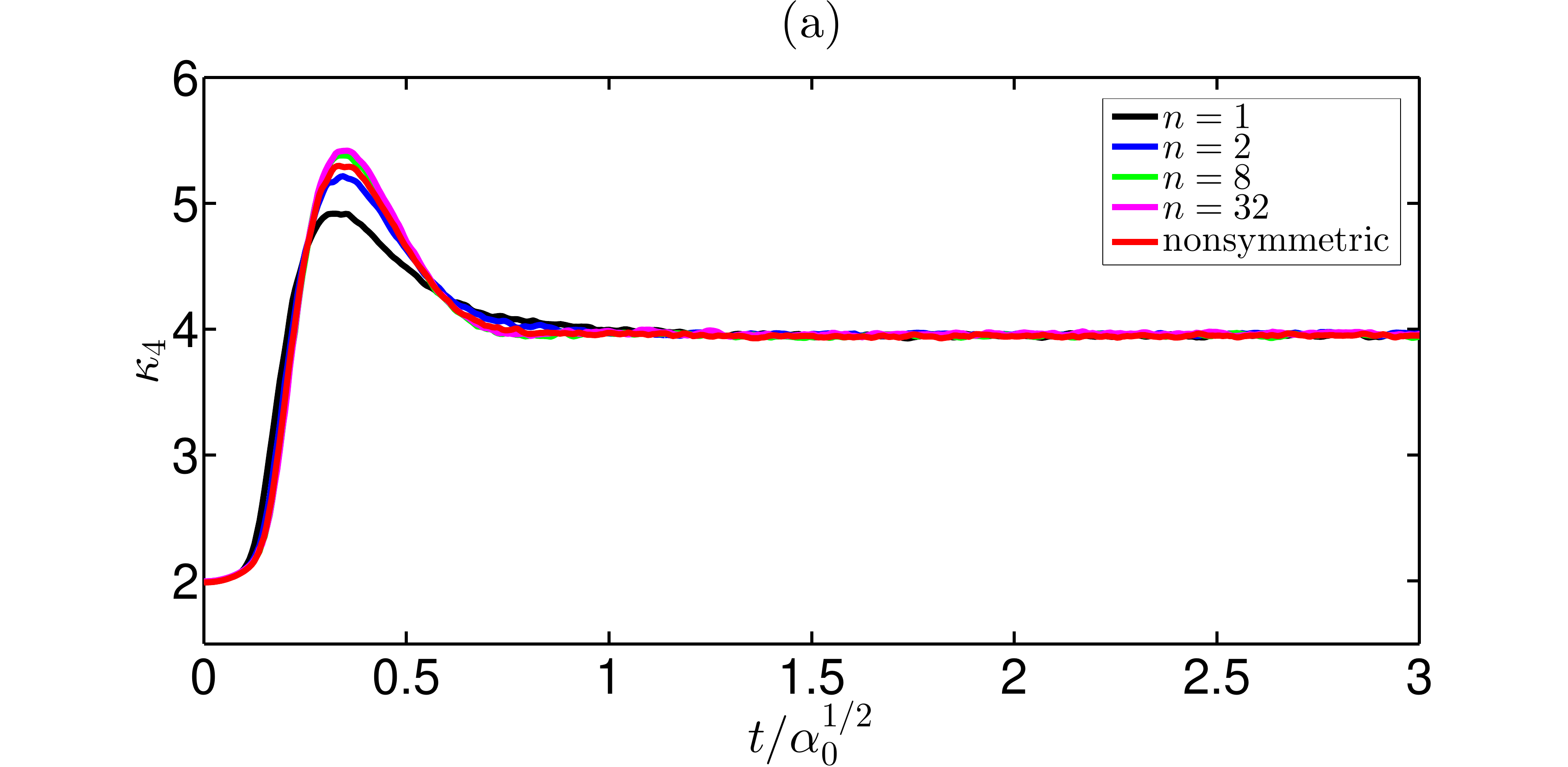}
\includegraphics[width=8.9cm]{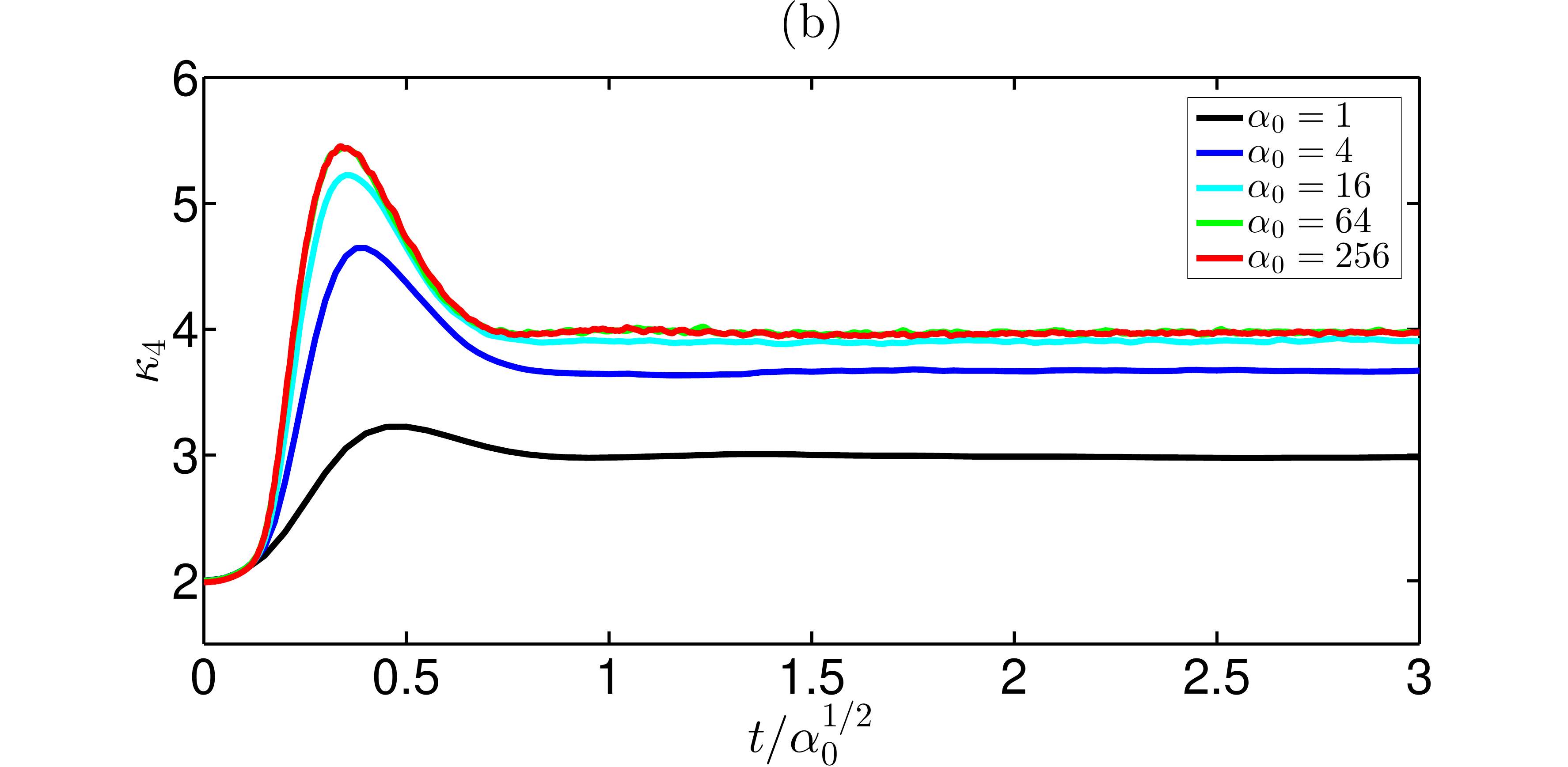}

\caption{\small {\it (Color on-line)} The fourth-order moment $\kappa_{4}$ vs. renormalized time $t/\sqrt{\alpha_{0}}$, for different numerical experiments. 
(a) Four simulations with ensembles of initial conditions having nonlinearity level $\alpha_{0}=64$ and different shapes of the spectrum (exponential $n=1$, Gaussian $n=2$, and super-Gaussian $n=8$ and $n=32$), and also one simulation with nonsymmetric initial spectrum of nonlinearity level $\alpha_{0}\approx 65.2$, see Appendix~\ref{Sec:AppA}. 
(b) Five simulations with initial conditions having different nonlinearity levels $\alpha_{0}=1$, $4$, $16$, $64$ and $256$; the shape of the initial spectrum is super-Gaussian with the exponent $n=32$. 
}
\label{fig:fig3}
\end{figure*}

We now continue with the description of a typical space-time dynamics for one realization of partially coherent wave of high nonlinearity and then describe the evolution of the (ensemble-averaged) statistical functions, on the example of numerical experiment with super-Gaussian initial spectrum~(\ref{IC-symmetric}) with the exponent $n=32$ and nonlinearity level $\alpha_{0}=64$. 

Fig.~\ref{fig:fig1}(a) illustrates one realization of a partially coherent wave; it looks as a collection of humps of characteristic width $\delta x\simeq 10$, that is in line with our estimate $\delta x\simeq \sqrt{\alpha_{0}}$ for $\alpha_{0}=64$. 
The spatiotemporal dynamics from this initial condition, shown in Fig.~\ref{fig:fig1}(b), reveals that shortly after beginning of the motion a breather-like structure (that -- as we know from the previous studies -- regularizes the gradient catastrophe) emerges on top of the largest hump. 
At $t\approx 1.35$ the structure reaches maximum amplitude of about three times larger than the initial one, and then disappears with the formation of two new breather-like structures on its left and right sides. 
The two structures then also reach maximum amplitudes and disappear with the emergence of three breather-like structures -- one between them and two other on their left and right sides. 
This process continues, and to time $t\simeq 5$ the breather-like structures fill the entire area of the largest hump. 
The smaller humps repeat the described scenario for the largest hump, but with longer characteristic times (compare the dynamics for the humps centered at $x\simeq -50$ and $x\simeq -70$ with that for $x=0$). 
Note that, even after sufficiently long evolution, the breather-like structures tend to stay within the initial area of the hump, that corroborates the suggestion of independence of the humps at the early stages of motion~\cite{tikan2019effect}. 
In the present paper, we concentrate on statistical characteristics after sufficiently long evolution (as we explain below, at $t\gtrsim 20$ for the experiment in Fig.~\ref{fig:fig1}); for description of the preceding dynamics we refer the reader to~\cite{bertola2013universality,randoux2017optical,tikan2017universality,tikan2019effect} where it was studied in more detail. 

The averaged over the ensemble of initial conditions kinetic $\langle H_{l}\rangle$ and potential $\langle H_{nl}\rangle$ energies, as well as the fourth-order moment $\kappa_{4}$, change rapidly until time $t\simeq 10$, see Fig.~\ref{fig:fig2}(a); below we call this rapid evolution the transient regime. 
After $t\simeq 10$, the three functions freeze, changing by less than $0.5$\% for $t\in[10,1000]$. 
From such a behavior, one might suggest that after $t\simeq 10$ the system arrives to the statistically steady state. 
If this suggestion is true, then the other statistical characteristics -- including the moments, the wave-action spectrum, the PDF and the autocorrelation of intensity -- must be independent of time for $t\gtrsim 10$. 

In Fig.~\ref{fig:fig2}(b-e) we compare these ensemble-averaged characteristics, additionally averaged over time intervals $t\in [20,40]$ (shown with black in the figures), $t\in [60,80]$ (blue), $t\in [180,200]$ (green) and $t\in [980,1000]$ (red). 
The time-averaging is applied since the corresponding functions evolve with time slowly, and this procedure allows to significantly improve the accuracy of our results most notably for the measurement of the PDF. 
As shown in the panels (b-d) of the figure, time-averaging in the intervals $t\in [20,40]$ and $t\in [60,80]$ gives almost identical results for the moments, wave-action spectrum and the PDF. 
Changes start to get visible from $t\simeq 200$, when the higher-order moments noticeably decrease, Fig.~\ref{fig:fig2}(b), the triangular shape of the wave-action spectrum at small wavenumbers $|k|\lesssim 1$ becomes less sharp and at large wavenumbers $|k|\gtrsim 10$ the spectrum decreases, Fig.~\ref{fig:fig2}(c), and the PDF visibly decreases at large intensities, Fig.~\ref{fig:fig2}(d). 

In contrast to the moments, wave-action spectrum and the PDF, the autocorrelation of intensity changes pronouncedly over the whole time interval $t\in[20,1000]$, as shown in Fig.~\ref{fig:fig2}(e). 
Its significant evolution with time is observed at intermediate distances $1\lesssim |x|\lesssim 20$ between the steady bell-shaped central peak of full width at half maximum $\Delta_{FWHM}\simeq 1.5$ and large distances $|x|\gtrsim 20$ where the autocorrelation reaches unity. 
Note that the distance, where the autocorrelation practically reaches unity, shrinks from $|x|\simeq 20$ at $t\simeq 40$ to $|x|\simeq 4$ at $t\simeq 1000$. 

Hence, after the transient regime, the system arrives not to the statistically stationary state, but to a \textit{quasi-stationary state} (QSS). 
In the QSS, most of the considered statistical functions change with time very slowly, and the evolution of statistics is hidden in the higher-order moments, the wave-action spectrum at large wavenumbers, the PDF at very large intensities and the autocorrelation of intensity at the intermediate distances. 

Note that for the experiments with smaller initial nonlinearity (e.g., for $\alpha_{0}=4$ and $16$) we observe faster convergence to the statistically steady state. 
For this reason, we believe that the statistically steady state exists for the experiments with larger initial nonlinearity (e.g., for $\alpha_{0}=64$ and $256$) as well, but the evolution towards it takes much longer time. 

Qualitatively the same scenario -- relatively short transient regime followed by the long QSS -- is observed for other numerical experiments with different initial spectra. 
Quantitatively, the transient regime is significantly affected by the profile and the nonlinearity level of the initial spectrum. 
This can be seen from evolution of the fourth-order moment $\kappa_{4}$ versus renormalized time $t/\sqrt{\alpha_{0}}$ shown in Fig.~\ref{fig:fig3}(a) for five experiments with different profiles of the initial spectrum (including the nonsymmetric spectrum constructed in Appendix~\ref{Sec:AppA}) and fixed nonlinearity level of the initial conditions, and also in Fig.~\ref{fig:fig3}(b) for five experiments with different nonlinearity levels and fixed profile of the initial spectrum.
Note that evolution of the ensemble-averaged kinetic and potential energies is defined uniquely from that of the fourth-order moment, as $\langle H_{nl}\rangle=-\kappa_{4}/2$ and $\langle H_{l}\rangle = -1 + 1/\alpha_{0} - \langle H_{nl}\rangle$. 
As follows from the figures, the duration of the transient regime $\delta T_{tr}$ is roughly equal to the square root of the initial nonlinearity, $\delta T_{tr}\simeq\sqrt{\alpha_{0}}$, and, during the transient, the fourth-order moment $\kappa_{4}$ takes different maximum values at slightly different renormalized times $t/\sqrt{\alpha_{0}}\simeq 0.35$ (compare with~\cite{tikan2019effect}), depending on the shape and the nonlinearity of the initial spectrum. 

From these results we can conclude that, while the duration of the transient regime increases with $\alpha_{0}$, the subsequent QSS lasts much longer than the transient before the final arrival to the statistically steady state. 
This means that, in the case of large initial nonlinearity $\alpha_{0}$, it is very difficult to reach the stationary state, both using the direct numerical simulations (DNS) and with an experiment in a real physical system. 
Indeed, examination of the stationary state with the DNS is difficult because the larger the initial nonlinearity $\alpha_{0}$, the larger (i) the required length of the simulation box $L$ (as the initial spectrum must be appropriately resolved, see Eqs.~(\ref{IC-ratio})-(\ref{IC-resolution})) and (ii) the duration of the transient regime; and then, the QSS takes a very long time as well. 
As for an experiment, the accumulated effect of the higher-order interactions over the long evolution should strongly affect the stationary state's statistics. 
For these reasons, in the present paper we concentrate on the detailed study of the statistical characteristics in the beginning of the QSS, which we perform in the next Section. 

Note also that, in the QSS, the fourth-order moment $\kappa_{4}$ practically does not change with time and its value is very close to $4$ for large enough initial nonlinearity $\alpha_{0}\ge 16$, see Fig.~\ref{fig:fig2}(a) and Fig.~\ref{fig:fig3}. 
For these reasons we think that, for large initial nonlinearity $\alpha_{0}\gg 1$, the asymptotic \textit{stationary} value of the fourth-order moment should also be very close to $4$. 
The latter determines the (asymptotic) ensemble-averaged potential energy, $\langle H_{nl}\rangle=-\kappa_{4}/2\approx -2$, as well as the kinetic one, $\langle H_{l}\rangle = -1 + 1/\alpha_{0} - \langle H_{nl}\rangle\approx 1$. 
Thus, the ratio of the potential energy to the kinetic one should be close to $2$ (compare with Fig.~\ref{fig:fig2}(a)) -- the same value as for the asymptotic stationary state of the noise-induced MI~\cite{agafontsev2015integrable}. 


\begin{figure*}[t]\centering
\includegraphics[width=8.9cm]{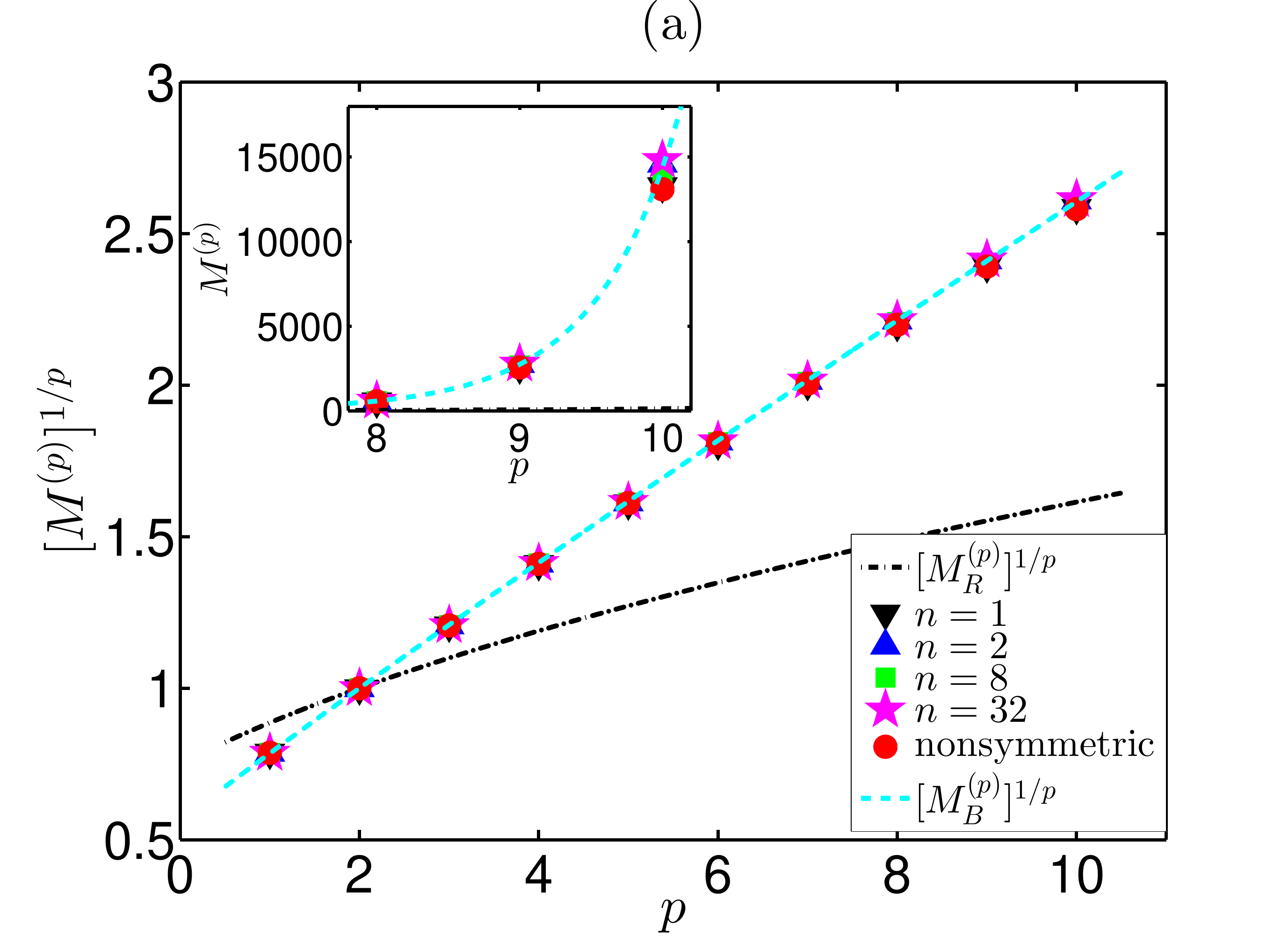}
\includegraphics[width=8.9cm]{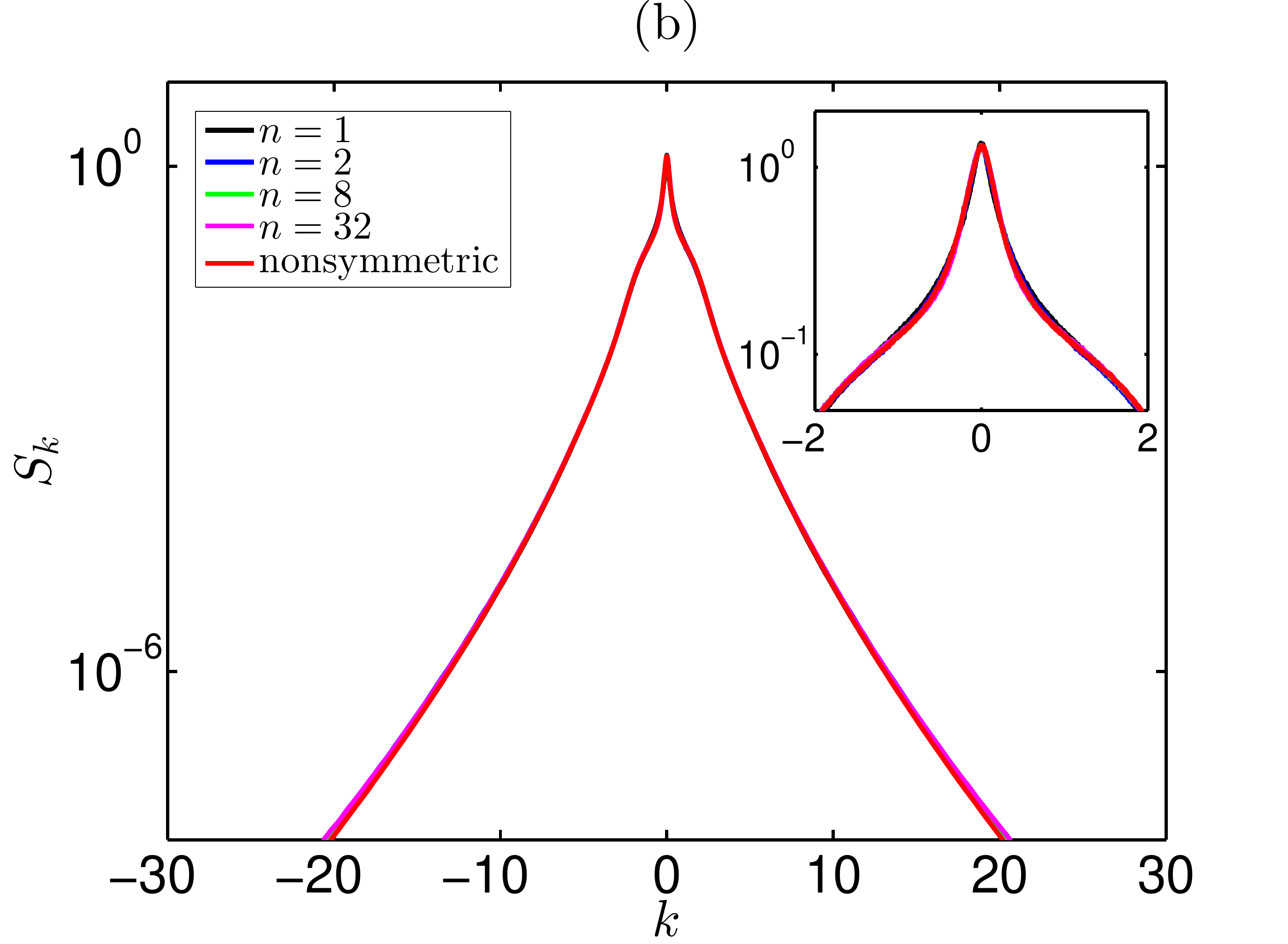}\\
\includegraphics[width=8.9cm]{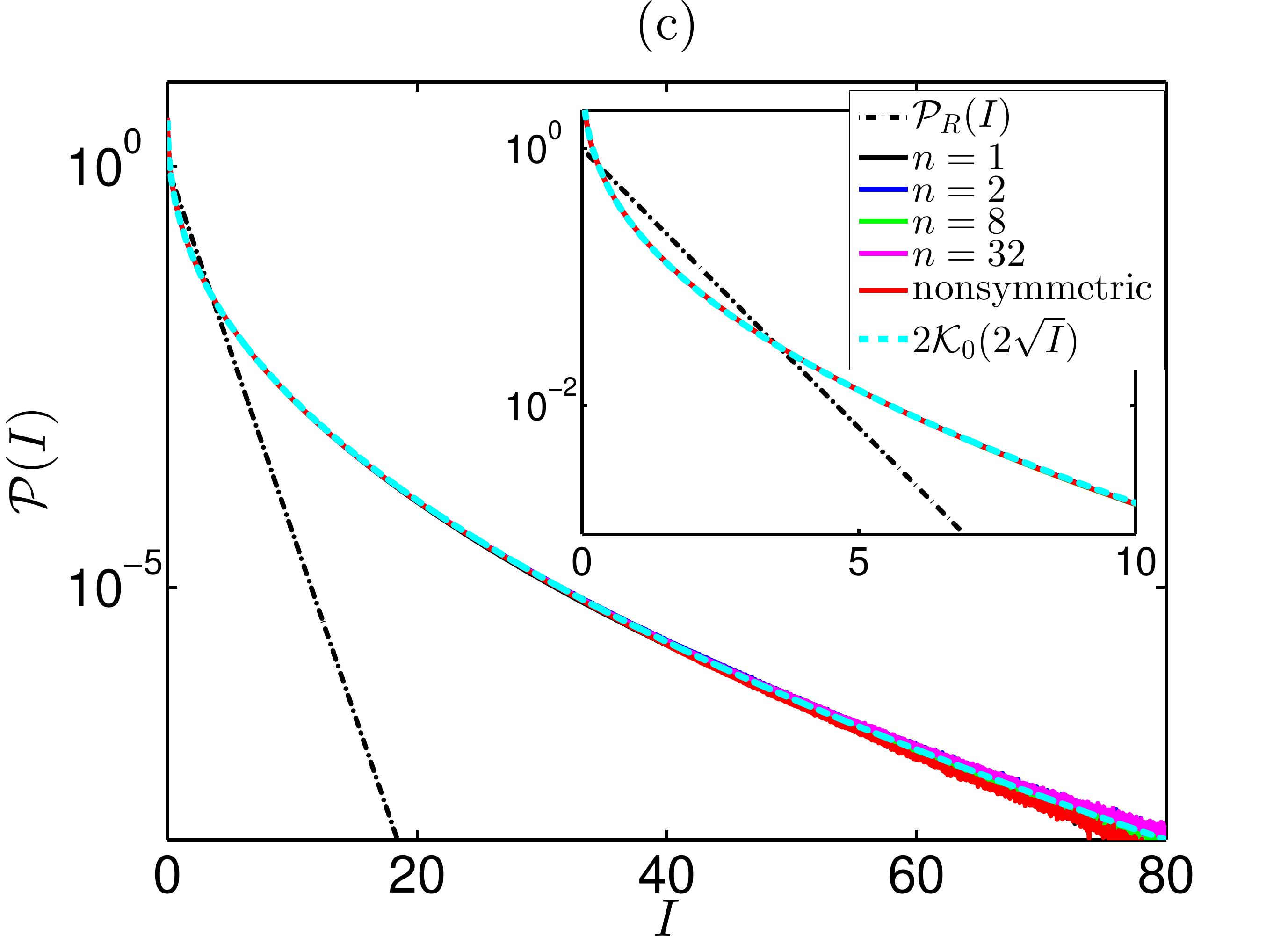}
\includegraphics[width=8.9cm]{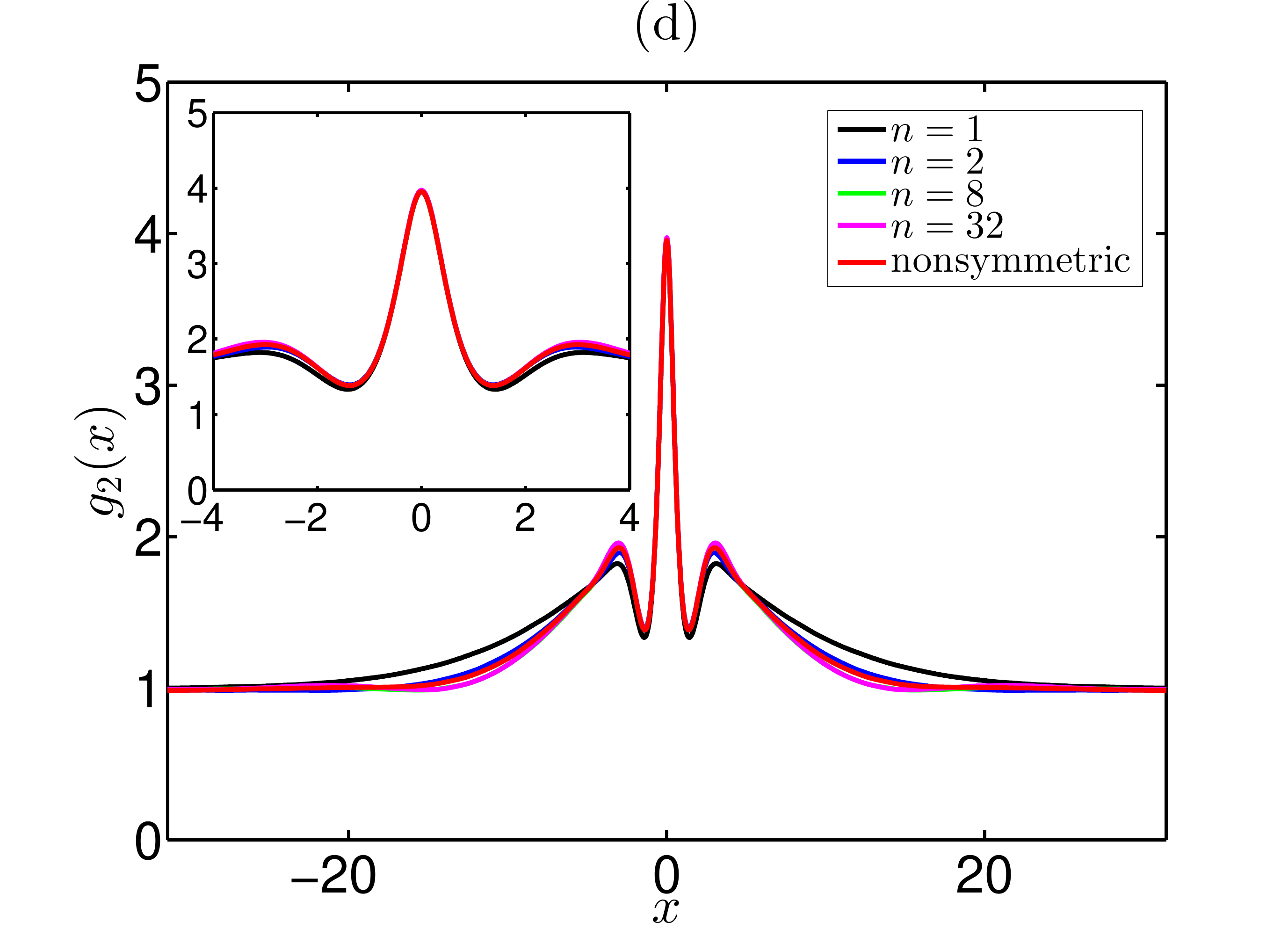}

\caption{\small {\it (Color on-line)} 
Ensemble- and time-averaged statistical characteristics in the beginning of the QSS for four super-Gaussian initial spectra with the exponents $n=1$, $2$, $8$, $32$ and one non-symmetric initial spectrum: (a) the moments $[M^{(p)}]^{1/p}$, (b) the wave-action spectrum $S_{k}$, (c) the PDF $\mathcal{P}(I)$ and (d) the autocorrelation of intensity $g_{2}(x)$. 
The initial nonlinearity is $\alpha_{0}=64$ for the four super-Gaussian spectra and $\alpha_{0}\approx 65.2$ for the non-symmetric spectrum, see Appendix~\ref{Sec:AppA}. 
The inset in the panel (a) shows the higher-order moments $M^{(p)}$ (without the $1/p$ power), and in the panels (b-d) -- the same functions as in the main figures with smaller scales. 
In the panels (c) and (a), the black dash-dot lines indicate the exponential PDF~(\ref{Rayleigh}) and the corresponding moments~(\ref{moments-Rayleigh}), while the cyan dashed lines -- the Bessel PDF~(\ref{PDF-Bessel}) and the corresponding moments~(\ref{moments-Bessel}).
}
\label{fig:fig4}
\end{figure*}

\begin{figure*}[t]\centering
\includegraphics[width=8.9cm]{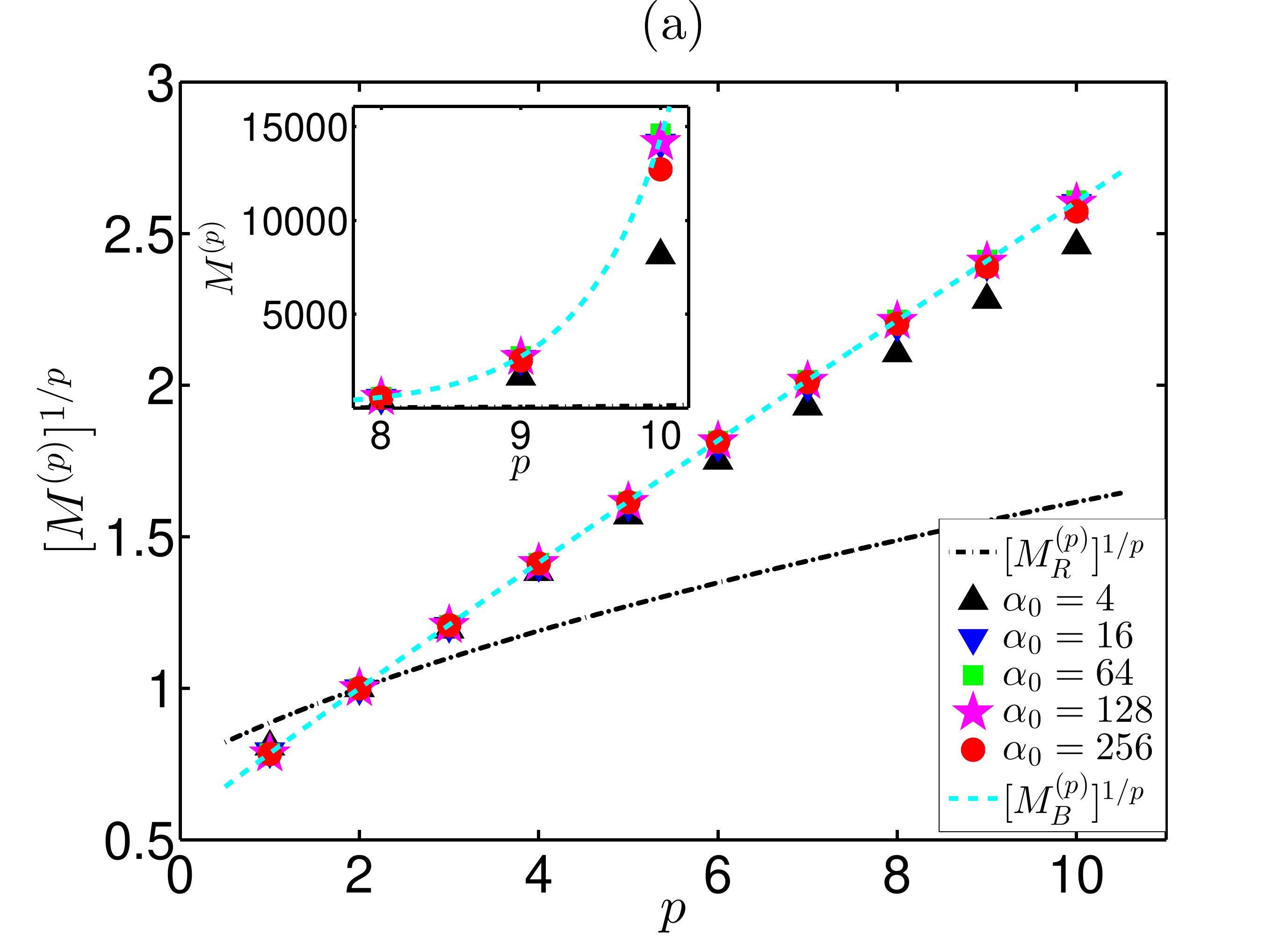}
\includegraphics[width=8.9cm]{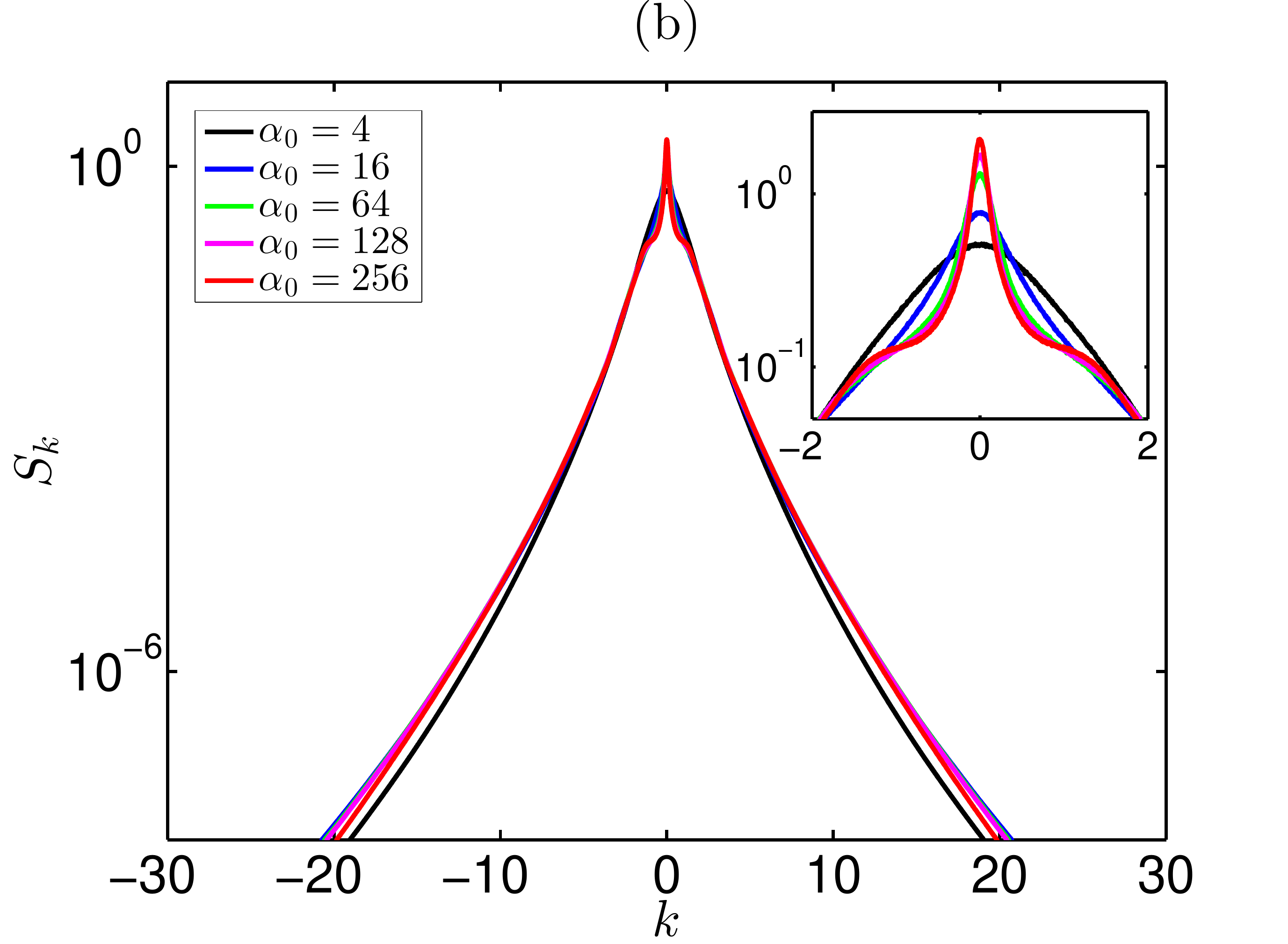}\\
\includegraphics[width=8.9cm]{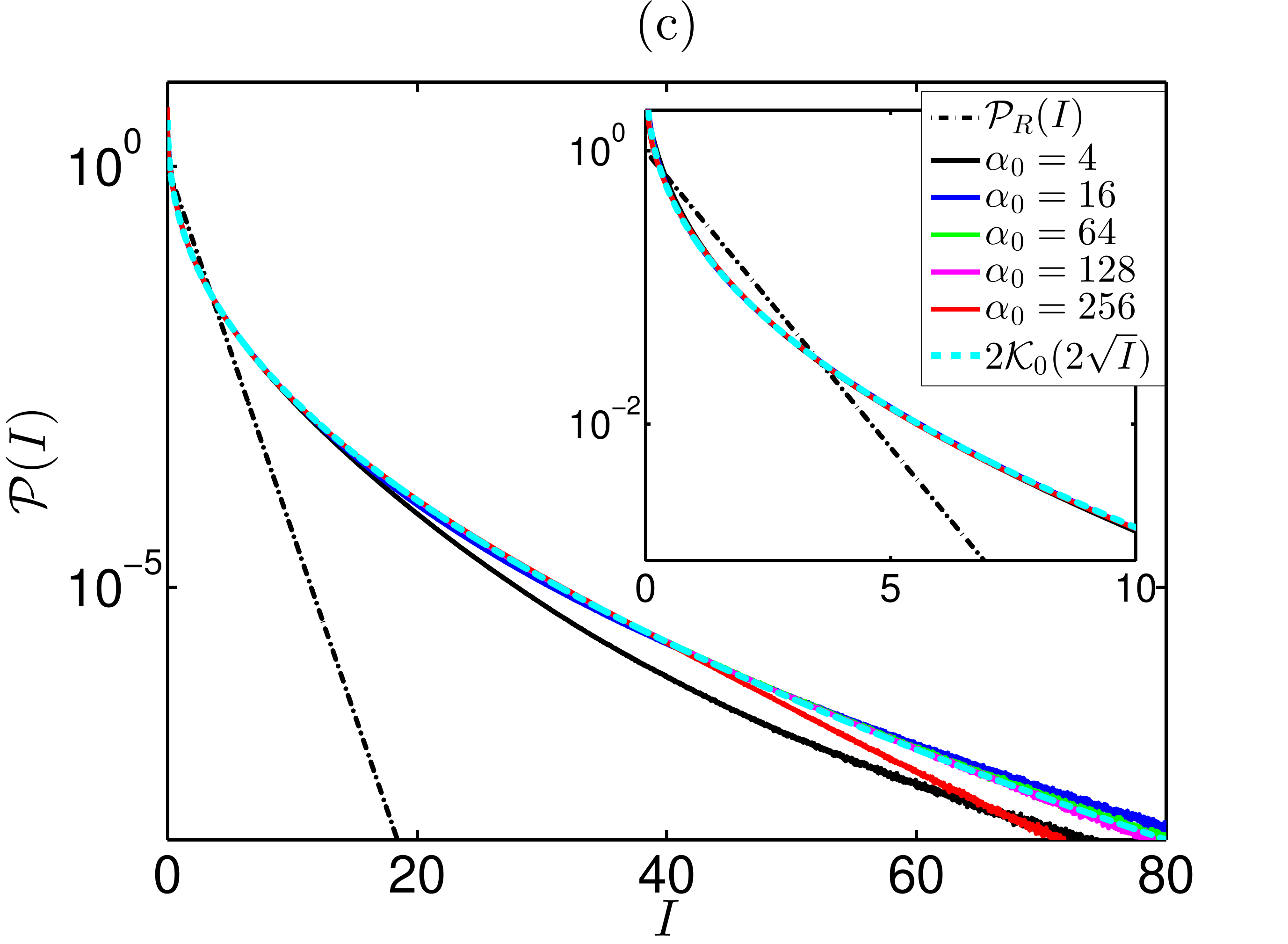}
\includegraphics[width=8.9cm]{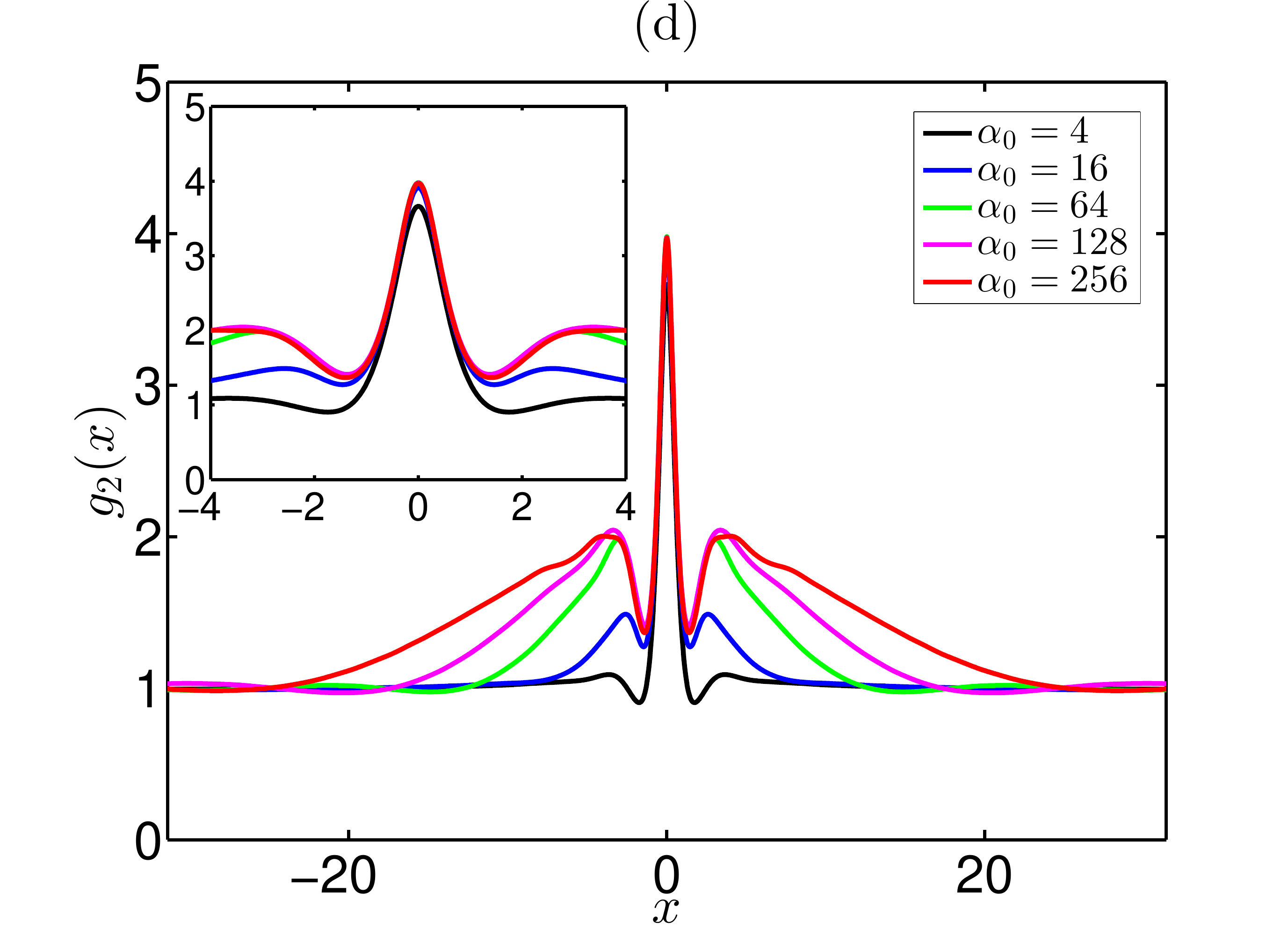}

\caption{\small {\it (Color on-line)} 
Ensemble- and time-averaged statistical characteristics in the beginning of the QSS for different initial nonlinearity levels $\alpha_{0}=4$, $16$, $64$, $128$ and $256$: (a) the moments $[M^{(p)}]^{1/p}$, (b) the wave-action spectrum $S_{k}$, (c) the PDF $\mathcal{P}(I)$ and (d) the autocorrelation of intensity $g_{2}(x)$. 
The initial spectrum is super-Gaussian with the exponent $n=32$. 
The inset in the panel (a) shows the higher-order moments $M^{(p)}$ (without the $1/p$ power), and in the panels (b-d) -- the same functions as in the main figures with smaller scales. 
In the panels (c) and (a), the black dash-dot lines indicate the exponential PDF~(\ref{Rayleigh}) and the corresponding moments~(\ref{moments-Rayleigh}), while the cyan dashed lines -- the Bessel PDF~(\ref{PDF-Bessel}) and the corresponding moments~(\ref{moments-Bessel}).
}
\label{fig:fig5}
\end{figure*}

\section{Statistics of the quasi-stationary state}
\label{Sec:Results2}

\subsection{Basic statistical functions}

To examine statistical characteristics of the QSS, we perform two sets of numerical experiments. 
In the first set, we fix the initial nonlinearity $\alpha_{0}$ and consider different profiles of the initial spectrum, including the nonsymmetric spectrum constructed in Appendix~\ref{Sec:AppA}. 
In the second set, we fix the shape of the initial spectrum and study different levels of $\alpha_{0}$. 
As in the previous Section, in addition to ensemble-averaging over random realizations of the initial conditions, we also perform time-averaging over relatively short time interval $t\in[t_{s},t_{e}]$ placed in the beginning of the QSS. 
The start of the interval is determined as $t_{s}/\sqrt{\alpha_{0}}=2.5$ in order to avoid the residual small oscillations of the fourth-order moment $\kappa_{4}$ visible for some experiments in Fig.~\ref{fig:fig3}(b). 
The duration of the interval is set the same for all experiments, $t_{e}-t_{s}=20$. 
We focus on examination of the kinetic and potential energies, the moments, the wave-action spectrum and the PDF, which change slowly during the QSS, and also provide results for the autocorrelation of intensity, which evolves pronouncedly at the intermediate distances.

Figures~\ref{fig:fig3}(a) and~\ref{fig:fig4} demonstrate statistical characteristics for fixed initial nonlinearity $\alpha_{0}=64$ and different profiles of the initial Fourier spectrum. 
We have checked that fixing $\alpha_{0}$ to other sufficiently large values leads to qualitatively the same results. 
As shown in the figures, in the beginning of the QSS, most of the considered statistical functions -- including the moments, the wave-action spectrum and the PDF of intensity -- practically do not depend on the profile of the initial spectrum, even when it is rather generic and nonsymmetric. 
In particular, the fourth-order moment $\kappa_{4}$ is very close to $4$, Fig.~\ref{fig:fig3}(a), and the moments $[M^{(p)}]^{1/p}$ increase with $p$ close to linearly, Fig.~\ref{fig:fig4}(a), exceeding significantly at large $p$ the moments~(\ref{moments-Rayleigh}) corresponding to the exponential PDF~(\ref{Rayleigh}). 

Note that we observe small deviations within $10\%$ for the absolute values of the tenth-order moment $M^{(10)}$, see the inset in Fig.~\ref{fig:fig4}(a). 
The difference of the same order is observed for $M^{(10)}$ when we repeat a numerical experiment for the second time -- that is, keeping the initial spectrum $A_{k}^{(0)}$ the same in Eq.~(\ref{IC}), but using a different random ensemble of the initial Fourier phases $\phi_{k}$. 
Additionally, the moments are directly connected to the PDF of intensity, and the difference in the latter in Fig.~\ref{fig:fig4}(c) is very small. 
For these reasons, we think that the deviations of the higher-order moments seen in the inset of Fig.~\ref{fig:fig4}(a) come mainly from the finiteness and randomness of the ensemble of the initial conditions and should diminish with the increase of the ensemble size. 
We disregard such deviations here, as well as below in similar cases. 

As shown in Fig.~\ref{fig:fig4}(b), the wave-action spectrum is symmetric, decays slightly slower than exponential at large wavenumbers $|k|\gtrsim 4$, and has a sharp triangular shape at small wavenumbers $|k|\lesssim 1$. 
The PDF is slightly smaller than the exponential distribution~(\ref{Rayleigh}) at moderate intensities $I\simeq 2$, and exceeds it by orders of magnitude at large intensities $I\gtrsim 5$, Fig.~\ref{fig:fig4}(c). 
In contrast to the wave-action spectrum and the PDF, the autocorrelation of intensity depends on the shape of the initial spectrum noticeably, and this dependency is observed at intermediate distances $2\lesssim |x|\lesssim 20$, Fig.~\ref{fig:fig4}(d), between the universal central peak of full width at half maximum $\Delta_{FWHM}\simeq 1.4$ and large distances $|x|\gtrsim 20$ where the autocorrelation reaches unity.
This means that the integrable system ``feels'' the shape of the initial spectrum even when it is very narrow. 

As we have checked, the other realizations of generic nonsymmetric initial spectrum characterized by the same value of $\alpha_{0}$ (see Appendix~\ref{Sec:AppA}) provide the same results for the moments, the wave-action spectrum and the PDF, while leading to noticeably different autocorrelation of intensity at the intermediate distances. 

The second set of numerical experiments with fixed super-Gaussian profile of the initial spectrum with the exponent $n=32$ and various nonlinearity levels $\alpha_{0}$ is shown in Fig.~\ref{fig:fig3}(b) and Fig.~\ref{fig:fig5}. 
In the beginning of the QSS, the fourth-order moment $\kappa_{4}$ increases with $\alpha_{0}$, from approximately $3$ for $\alpha_{0}=1$ to very close to $4$ for $\alpha_{0}\ge 16$, see Fig.~\ref{fig:fig3}(b). 
The moments $[M^{(p)}]^{1/p}$ turn out to be practically a universal function increasing almost linearly with $p$ for nonlinearities $16\le\alpha_{0}\le 128$, Fig.~\ref{fig:fig5}(a). 
When the initial nonlinearity is not large enough, $\alpha_{0}\le 4$, the higher-order moments are noticeably smaller than this function. 
For the largest nonlinearity $\alpha_{0}=256$ that we study in the present paper, we also detect slightly smaller values for the higher-order moments. 
In particular, the tenth-order moment $M^{(10)}$ is more than $40\%$ smaller for $\alpha_{0}=4$ and about $10\%$ smaller for $\alpha_{0}=256$ than those for $16\le\alpha_{0}\le 128$, as shown in the inset of Fig.~\ref{fig:fig5}(a). 

The wave-action spectrum depends significantly on $\alpha_{0}$, Fig.~\ref{fig:fig5}(b), that is most noticeable at small wavenumbers $|k|\lesssim 1$, where the spectrum forms a triangular sharpening with increasing $\alpha_{0}$. 
For sufficiently large $\alpha_{0}$, the spectrum at these wavenumbers resembles the asymptotic spectrum of the noise-induced MI~\cite{agafontsev2015integrable}.
The PDF for $16\le\alpha_{0}\le 128$ turns out to be practically a universal function, which exceeds the exponential distribution~(\ref{Rayleigh}) by orders of magnitude at large intensities $I\gtrsim 5$, Fig.~\ref{fig:fig5}(c). 
For smaller $\alpha_{0}\le 4$ and larger $\alpha_{0}\ge 256$ initial nonlinearities, we observe the PDF slightly smaller at large intensities, with deviations by less than one order of magnitude within the interval $I\in[20, 80]$. 
These deviations explain the smaller values for the higher-order moments compared to $16\le\alpha_{0}\le 128$, and we will address them in more detail in the next paragraphs. 
The autocorrelation of intensity $g_{2}(x)$ is almost a universal bell-shaped function at small distances $|x|\lesssim 1.4$, while changing significantly with $\alpha_{0}$ at the intermediate distances where it approaches to unity, Fig.~\ref{fig:fig5}(d). 
The distance where the autocorrelation practically reaches unity increases with the initial nonlinearity roughly as $|x|\propto\sqrt{\alpha_{0}}$, as can be seen in the figure (compare $g_{2}(x)$ for $\alpha_{0}=4$, $16$, $64$ and $256$).
Note that for $\alpha_{0}=4$ the maximum of the autocorrelation function is slightly smaller than $4$, that reflects the smaller value of the fourth-order moment $\kappa_{4}$ in the QSS for this experiment, see Fig.~\ref{fig:fig3}(b). 

\begin{figure*}[t]\centering
\includegraphics[width=17.9cm]{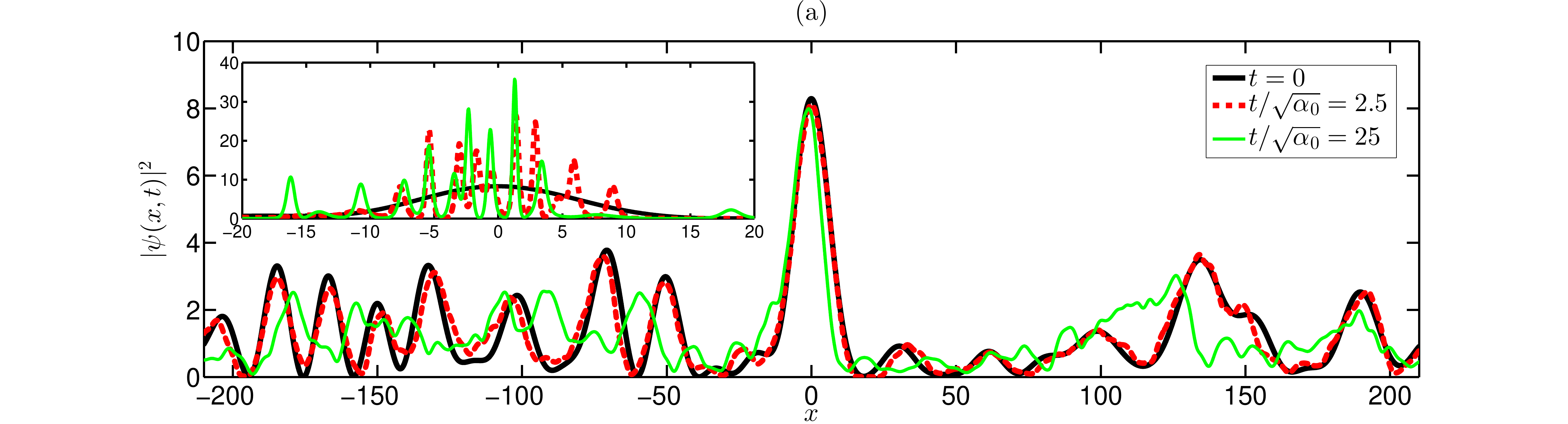}\\
\includegraphics[width=17.9cm]{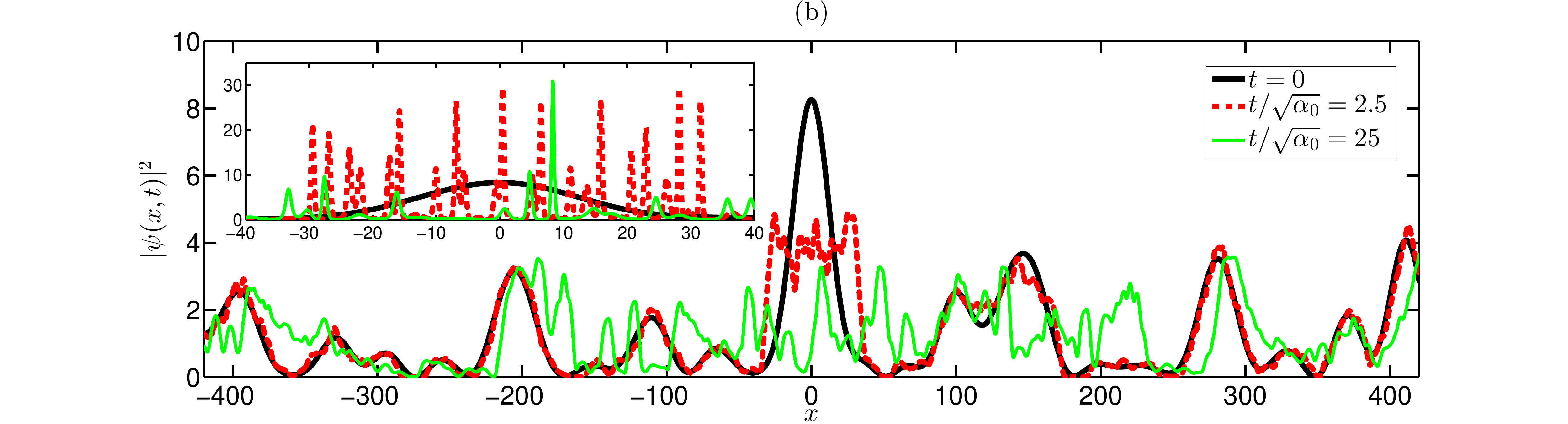}

\caption{\small {\it (Color on-line)} Smoothed intensity $|\psi|^{2}$ versus spatial coordinate $x$ at different times for one realization of the initial conditions, for the initial nonlinearity (a) $\alpha_{0}=64$ (the same realization as in Fig.~\ref{fig:fig1}) and (b) $\alpha_{0}=256$; the initial spectrum is super-Gaussian with the exponent $n=32$.
The black lines indicate the initial (non-smoothed) intensity at $t=0$, the dashed red and thin green lines -- the smoothed intensity at renormalized times $t/\sqrt{\alpha_{0}}=2.5$ and $t/\sqrt{\alpha_{0}}=25$, respectively. 
The time $t/\sqrt{\alpha_{0}}=2.5$ corresponds to the beginning of the QSS. 
The spatial smoothing is performed with the weighted local regression (lowess) filter over the window $\ell=4\pi$. 
The insets show the original (non-smoothed) intensity at the same times on the scale of the largest hump; note the different vertical scales compared to the main figures. 
The maximum of the initial amplitude is shifted to $x=0$ for better visualization. 
}
\label{fig:fig6}
\end{figure*}

\begin{figure*}[t]\centering
\includegraphics[width=8.9cm]{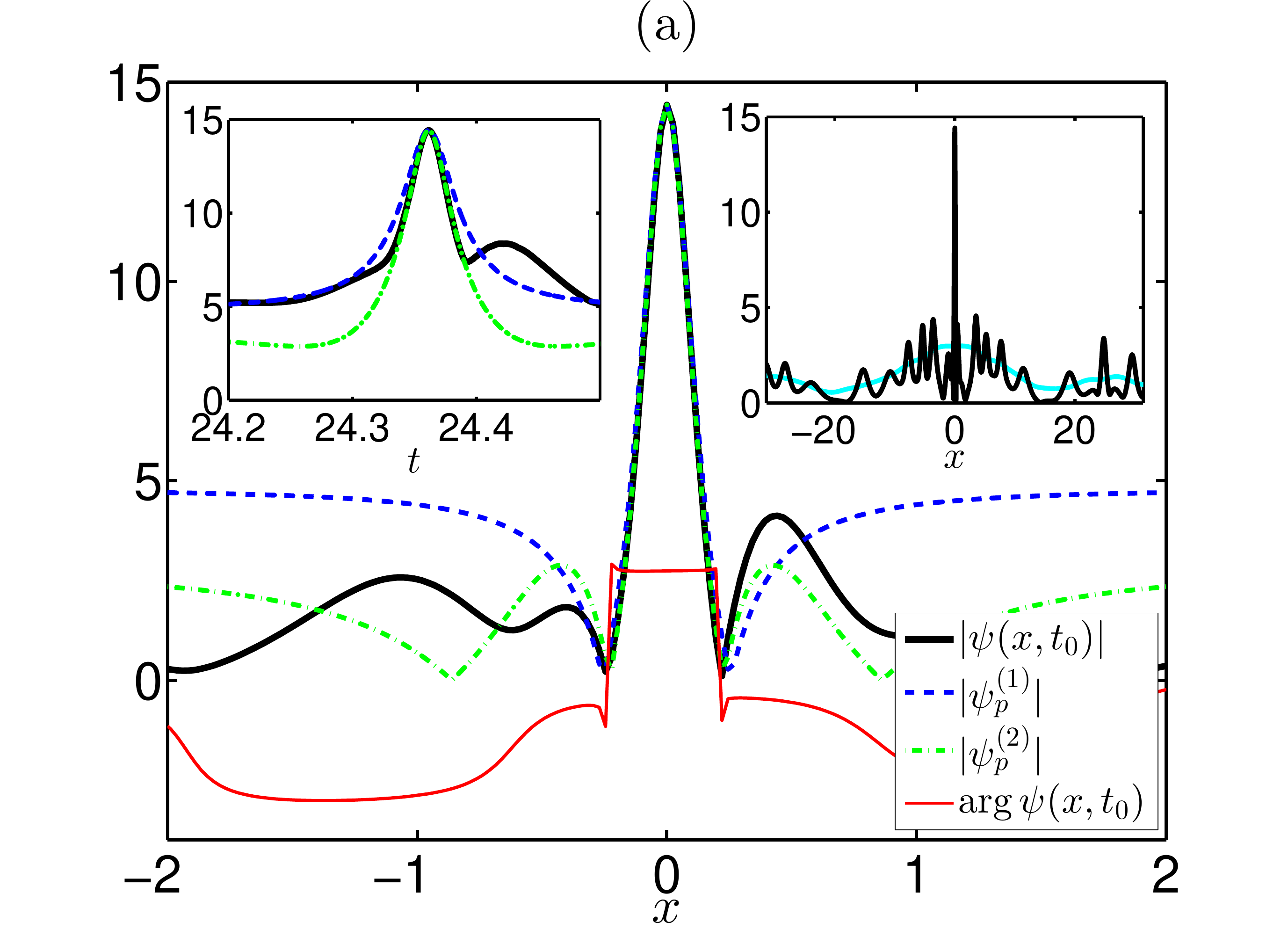}
\includegraphics[width=8.9cm]{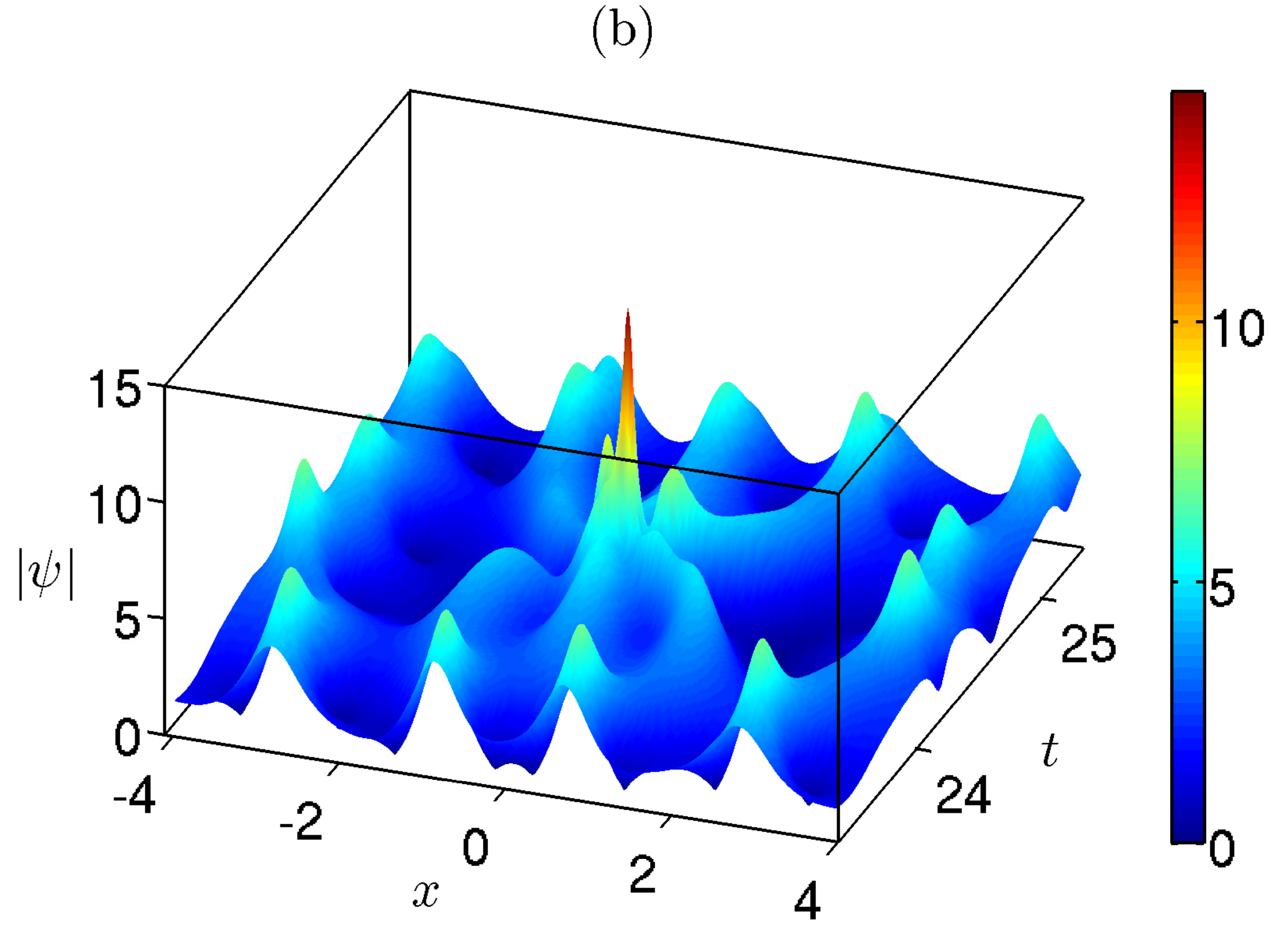}

\caption{\small {\it (Color on-line)} 
The largest rogue wave detected in the QSS for the ensemble with super-Gaussian initial spectrum with the exponent $n=32$ and nonlinearity level $\alpha_{0}=64$; the maximum amplitude is shifted to $x=0$ for better visualization. 
(a) The thick black and the thin red lines show the coordinate dependencies for the amplitude $|\psi|$ and the phase $\mathrm{arg}\,\psi$, while the dashed blue and the dash-dot green indicate fits by the rational breather solutions of the first and the second orders. 
The left inset demonstrates time-dependency for the maximum amplitude $\max_{x}|\psi|$ (thick black) and its fits with the corresponding rational breather solutions (dashed blue and dash-dot green), while the right inset shows the amplitude $|\psi|$ (black) versus the smoothed amplitude (cyan) in the $x$-space at the time of the maximum elevation. 
The smoothing is performed with the lowess filter over the window $\ell=4\pi$. 
(b) Space-time representation of the amplitude $|\psi(x,t)|$ near the rogue wave event. 
}
\label{fig:fig7}
\end{figure*}

\subsection{Insights from wavefield dynamics}

The deviations of the PDF for $\alpha_{0}\le 4$ and $\alpha_{0}\ge 256$ from almost a universal function observed for $16\le\alpha_{0}\le 128$ can be understood qualitatively by looking at the wavefield dynamics. 
Each realization of the initial conditions represents a collection of humps of characteristic width $\delta x\simeq\sqrt{\alpha_{0}}$, and the smaller-scale breather-like structures of width $\simeq 1$ gradually emerge on each hump at the early stage of the evolution, see e.g. Fig.~\ref{fig:fig1} and~\cite{bertola2013universality,randoux2017optical,tikan2017universality,tikan2019effect}. 
Then, if we smooth the intensity over the spatial dimension using some intermediate-length window eliminating these smaller-scale structures, we might be able to observe qualitatively the evolution of the humps -- for instance, how quickly they split to smaller ones and mix with other humps. 
Figure~\ref{fig:fig6} demonstrates the results of this smoothing performed with the weighted local regression (lowess) filter~\cite{cleveland1988regression} over the spatial window $\ell=4\pi$ in comparison with the initial intensity, for one realization of the initial conditions with nonlinearity $\alpha_{0}=64$ (Fig.~\ref{fig:fig6}(a)) and $\alpha_{0}=256$ (Fig.~\ref{fig:fig6}(b)). 
Note that, as we have checked, the smoothing windows from $\ell=\pi$ to $8\pi$ provide qualitatively the same results. 
The smoothed intensity is shown in the figures at two different times $t/\sqrt{\alpha_{0}}=2.5$ and $t/\sqrt{\alpha_{0}}=25$, the first of which corresponds to the beginning of the QSS; the insets of the figures show the original (non-smoothed) intensity at the same times on the scale of the largest hump. 

For the first experiment with $\alpha_{0}=64$, one can see that, while the original (non-smoothed) intensity in the beginning of the QSS may exceed the maximum of the initial intensity by more than $3$ times due to the emergence of the smaller-scale breather-like structures, the smoothed intensity at the same time practically coincides with the initial one, Fig.~\ref{fig:fig6}(a). 
During the subsequent evolution, the humps revealed by the smoothing procedure gradually mix with each other and disappear, though this turns out to be a long process. 
In particular, the largest hump with the maximum intensity $|\psi|^2\simeq 8$ survives and retains its shape even at $t/\sqrt{\alpha_{0}}=25$, that corresponds to $t=200$, i.e., the time-averaging interval $t\in[180,200]$ shown in Fig.~\ref{fig:fig2}(b-e) with the green color. 
Qualitatively the same scenario is observed for other realizations of the initial conditions with the initial nonlinearity from $\alpha_{0}=16$ to $128$. 

The above observations allow to suggest the analytical form of the universal PDF for $16\le\alpha_{0}\le 128$. 
Indeed, in the QSS the smoothed intensity $I_{s}$ evolves slowly, and in the beginning of this state it practically coincides with the initial intensity. 
The latter is exponentially distributed, so that in the beginning of the QSS the smoothed intensity is exponentially distributed too, $\mathcal{P}_{s}(I_{s})=\exp(-I_{s})$. 
Let us consider a sufficiently small element of a hump with smoothed intensity $I_{s}$. 
This element provides a contribution $\mathcal{P}_{e}(I, I_{s})$ to the overall PDF of intensity $I=|\psi|^{2}$ (non-smoothed), depending on the smaller-scale breather-like structures evolving on its background. 
Assuming the simplest scenario that each such contribution is exponential distribution with the mean intensity $I_{s}$, i.e., $\mathcal{P}_{e}(I, I_{s})=I_{s}^{-1}\exp(-I/I_{s})$, we come to the following estimate for the overall PDF,
\begin{eqnarray}
&&\mathcal{P}_{B}(I) = \int_{0}^{+\infty}\mathcal{P}_{e}(I, I_{s})\times\mathcal{P}_{s}(I_{s})\,dI_{s} = \nonumber\\
&& =\int_{0}^{+\infty}I_{s}^{-1}e^{-I_{s}-I/I_{s}}\,dI_{s} = 2\mathcal{K}_{0}(2\sqrt{I}), \label{PDF-Bessel}
\end{eqnarray}
where $\mathcal{K}_{0}$ denotes the modified Bessel function of the second kind of zeroth order. 
The estimate~(\ref{PDF-Bessel}) is shown in Fig.~\ref{fig:fig4}(c) and Fig.~\ref{fig:fig5}(c) with the dashed cyan lines, and for the initial nonlinearities $16\le\alpha_{0}\le 128$ it is in a remarkable agreement with the numerical simulations. 

The Bessel PDF~(\ref{PDF-Bessel}) represents a strongly non-exponential distribution which decays at large intensity as
$$
\mathcal{P}_{B}(I) \to \frac{\sqrt{\pi}}{I^{1/4}}\,e^{-2\sqrt{I}}\quad\mathrm{for}\quad I\to +\infty,
$$
and corresponds to the moments 
\begin{equation}\label{moments-Bessel}
M^{(p)}_{B} = \int_{0}^{+\infty}|\psi|^{p}\,\mathcal{P}_{B}(|\psi|^{2})\,d|\psi|^{2} = \Gamma_{1+p/2}^{2},
\end{equation}
in particular, the fourth-order moment $\kappa_{4}=4$. 
Note that, quite remarkably, the latter values equal to the squares of the moments~(\ref{moments-Rayleigh}) corresponding to the exponential PDF~(\ref{Rayleigh}). 
The moments following from the Bessel PDF are in a very good agreement with those observed numerically in the beginning of the QSS, as can be seen in Fig.~\ref{fig:fig4}(a) and Fig.~\ref{fig:fig5}(a) for $16\le\alpha_{0}\le 128$. 
Then, the close to linear dependency of $[M^{(p)}]^{1/p}$ with $p$ can be explained by the Stirling's formula applied to Eq.~(\ref{moments-Bessel}). 

For the experiment shown in Fig.~\ref{fig:fig6}(b) with the initial nonlinearity $\alpha_{0}=256$, the behavior of the smoothed intensity has a significant difference. 
The largest hump in this case substantially decreases in intensity and broadens already to the beginning of the QSS, and then splits to several smaller humps. 
Note that this process can be seen even by looking at the original (non-smoothed) intensity, as indicated in the inset of the figure. 
The smaller humps, however, may preserve their shape for a sufficiently long time. 
We think that the splitting of the largest humps, which are the background for the evolving on them smaller-scale breather-like structures, is the reason why the PDF for $\alpha_{0}=256$ is slightly smaller at large intensities $I\gtrsim 50$ than the estimate~(\ref{PDF-Bessel}). 

Performing experiments with individual realizations characterized by even larger initial nonlinearity $\alpha_{0}>256$, we observe that, in the beginning of the QSS, the tendency for splitting of the humps depends on the humps' width and maximum intensity. 
Specifically, for fixed width (which is proportional to $\sqrt{\alpha_{0}}$), humps with maximum intensity larger than some threshold tend to substantially decrease in intensity and broaden, and then split to smaller humps, while the initially smaller humps may survive for a long time. 
The value of the intensity threshold, dividing the different behavior, decreases with increasing $\alpha_{0}$. 
Hence, we expect that, for larger nonlinearity $\alpha_{0}>256$, the PDF should deviate from the Bessel estimate~(\ref{PDF-Bessel}) even more pronouncedly and starting from smaller intensities. 
This should result in the higher-order moments smaller than the estimate~(\ref{moments-Bessel}) too, with the deviation increasing with $\alpha_{0}$. 

Note that, when the initial nonlinearity is not large enough, $\alpha_{0}\le 4$, the humps have characteristic width of unity order, $\delta x\lesssim 2$. 
Then, the decomposition of evolution by those of the humps and of the breather-like structures should be inapplicable, as the latter have width of unity order too, and it is natural to expect the PDF deviating from the Bessel estimate~(\ref{PDF-Bessel}). 
We indeed observe this deviation in the numerical simulation, as demonstrated in Fig.~\ref{fig:fig5}(c). 

Summarizing the experiments shown in Fig.~\ref{fig:fig6}, we conclude that, in the QSS, the evolution of the wavefield can be subdivided by the fast changes with time of the smaller-scale breather-like structures moving on the background of the humps and the slow mixing and disappearance of the humps. 
For the initial nonlinearity $16\le\alpha_{0}\le 128$, all humps evolve slowly. 
For larger values $\alpha_{0}\ge 256$, the largest humps tend to significantly decrease in intensity and broaden already to the beginning of the QSS, and then split to several smaller humps, while the initially smaller humps survive and retain their shape for a long time. 
We believe that the slow evolution of the humps is the main process that underlies the QSS and determines the slow changes with time of its basic statistical characteristics. 
In our opinion, in the asymptotic \textit{stationary} state all humps should disappear, so that the smoothed intensity should everywhere be sufficiently close to unity. 

\subsection{Rogue waves}

Figure~\ref{fig:fig7} shows the largest rogue wave event detected in the QSS for the experiment with super-Gaussian initial spectrum with the exponent $n=32$ and nonlinearity $\alpha_{0}=64$. 
This event occurred at time $t_{0}\approx 24.36$, had duration $\Delta T\simeq 0.07$ and reached maximum amplitude $\max|\psi|\approx 14.4$ corresponding to intensity $I\approx 207$. 
The rogue wave emerged on the background of the hump having maximum amplitude close to $3$ and exceeded the latter by about $5$ times, see the right inset in Fig.~\ref{fig:fig7}(a). 
We observe such behavior for other rogue wave events as well: the initial conditions contain large humps exceeding the average amplitude by several times, and the evolution of the smaller-scale breather-like structures on the background of the humps leads to spikes exceeding the amplitude of the humps by several times more. 

For the experiment with super-Gaussian initial spectrum with the exponent $n=32$ and nonlinearity $\alpha_{0}=64$, the maximum of the fourth-order moment is reached at $t\approx 2.8$, see Fig.~\ref{fig:fig3}(a), so that the rogue wave demonstrated in Fig.~\ref{fig:fig7} is not the Peregrine-like coherent structure regularizing the gradient catastrophe in the transient regime. 
Nevertheless, its spatial profile $|\psi(x,t_{0})|$ at the time of its maximal elevation $t_{0}$, as well as the temporal evolution of the maximum amplitude $\max_{x}|\psi|$, are very well approximated by the so-called rational breather solution of the second order, as shown in Fig.~\ref{fig:fig7}(a). 

The rational breather solution of the first order, also known as the Peregrine breather~\cite{peregrine1983water}, is localized in space and time rational solution of the 1D-NLSE~(\ref{NLSE}), 
\begin{equation}\label{Peregrine1}
\psi_{p}^{(1)}(x,t)=e^{i\,t}\bigg[1-\frac{4(1+2 i t)}{1+2 x^{2}+4 t^{2}}\bigg].
\end{equation}
The next-order rational breathers are too cumbersome, and we refer the reader to~\cite{akhmediev2009rogue} where they were first found. 
If $\psi_{p}(x,t)$ is a solution of the 1D-NLSE, then $A_{0}\psi_{p}(\varkappa,\sigma)$, where $\varkappa = |A_{0}|(x-x_{0})$ and $\sigma = |A_{0}|^{2}(t-t_{0})$, is also a solution. 
In Fig.~\ref{fig:fig7}(a), the rational breathers of the first (dashed blue) and the second (dash-dot green) orders are scaled with parameters $A_{0}$, $x_{0}$ and $t_{0}$ to fit the observed rogue wave in its maximum amplitude, position and time of occurrence. 
Note that the phase of the rogue wave $\mathrm{arg}\,\psi(x,t_{0})$ is almost constant near the amplitude maximum, as in the case of the rational breathers at the time of their maximal elevation. 
The space-time representation of the rogue wave shown in Fig.~\ref{fig:fig7}(b) looks like a collision of three pulses with width of unity order. 

For each of our numerical experiments with initial nonlinearity from $\alpha_{0}=4$ to $256$ and with different profiles of the initial spectrum including the nonsymmetric spectra, we have checked the largest $20$ rogue waves detected in the beginning of the QSS.
We have found that all of these rogue waves are very well approximated by the rational breather solutions of either the first (the Peregrine breather), or the second orders, with the flat phase profile $\mathrm{arg}\,\psi(x,t_{0})\approx\mathrm{const}$ near the amplitude maximum. 


\section{Effects of additional wide-spectrum noise}
\label{Sec:Results3}

As we have mentioned in Section~\ref{Sec:Formulation}, in physical systems high nonlinearity can be achieved by both increasing the intensity and decreasing the spectral width. 
However, increasing the intensity above certain limits may not be desirable because of the increased influence of the higher-order effects beyond the 1D-NLSE. 
There are no such limitations for decreasing of the spectral width. 
Nevertheless, real physical systems have noise and its effect may be significant. 
In this Section, we demonstrate two numerical experiments with inclusion of additional small and moderate wide-spectrum noise and discuss the effects on the statistical results. 

\begin{figure*}[t]\centering
\includegraphics[width=11.9cm]{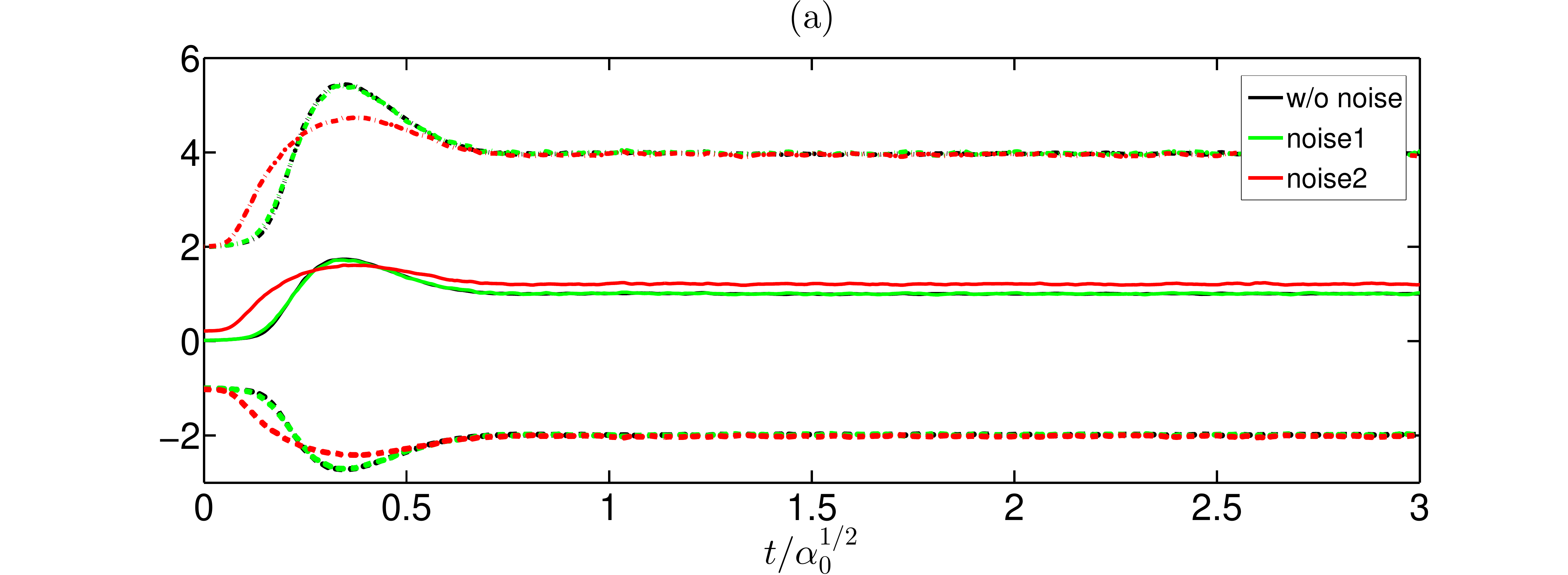}
\includegraphics[width=5.9cm]{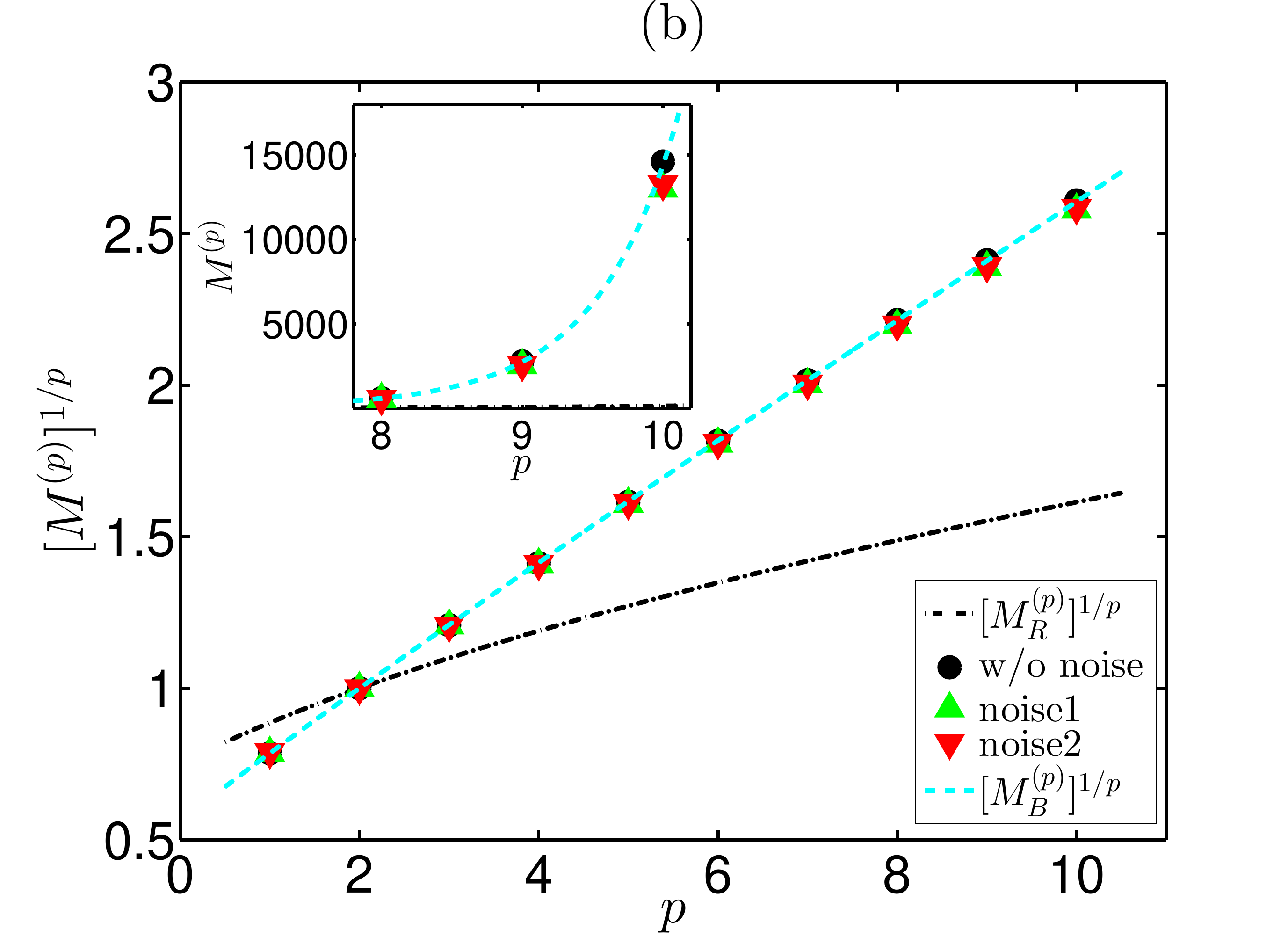}\\
\includegraphics[width=5.9cm]{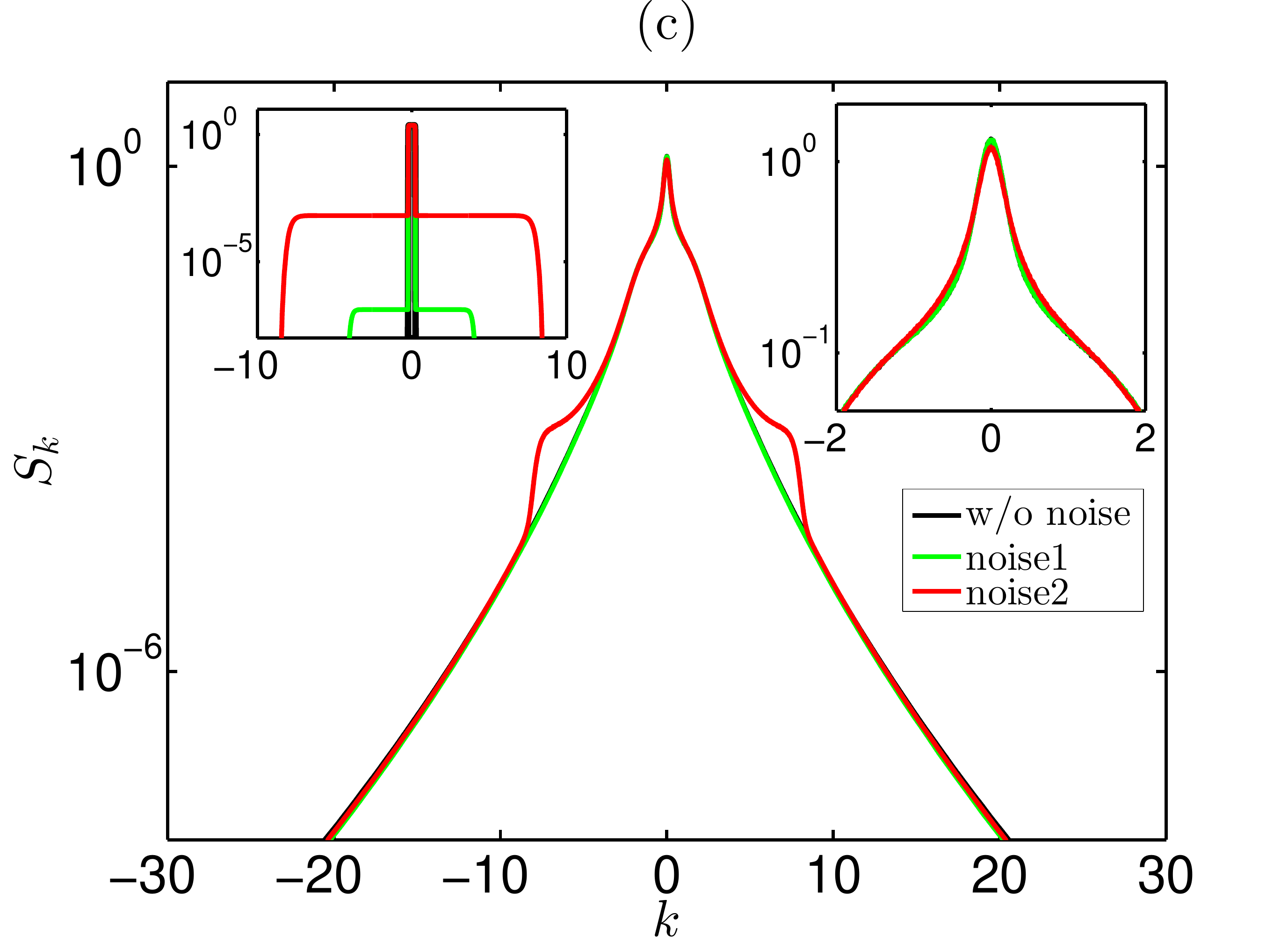}
\includegraphics[width=5.9cm]{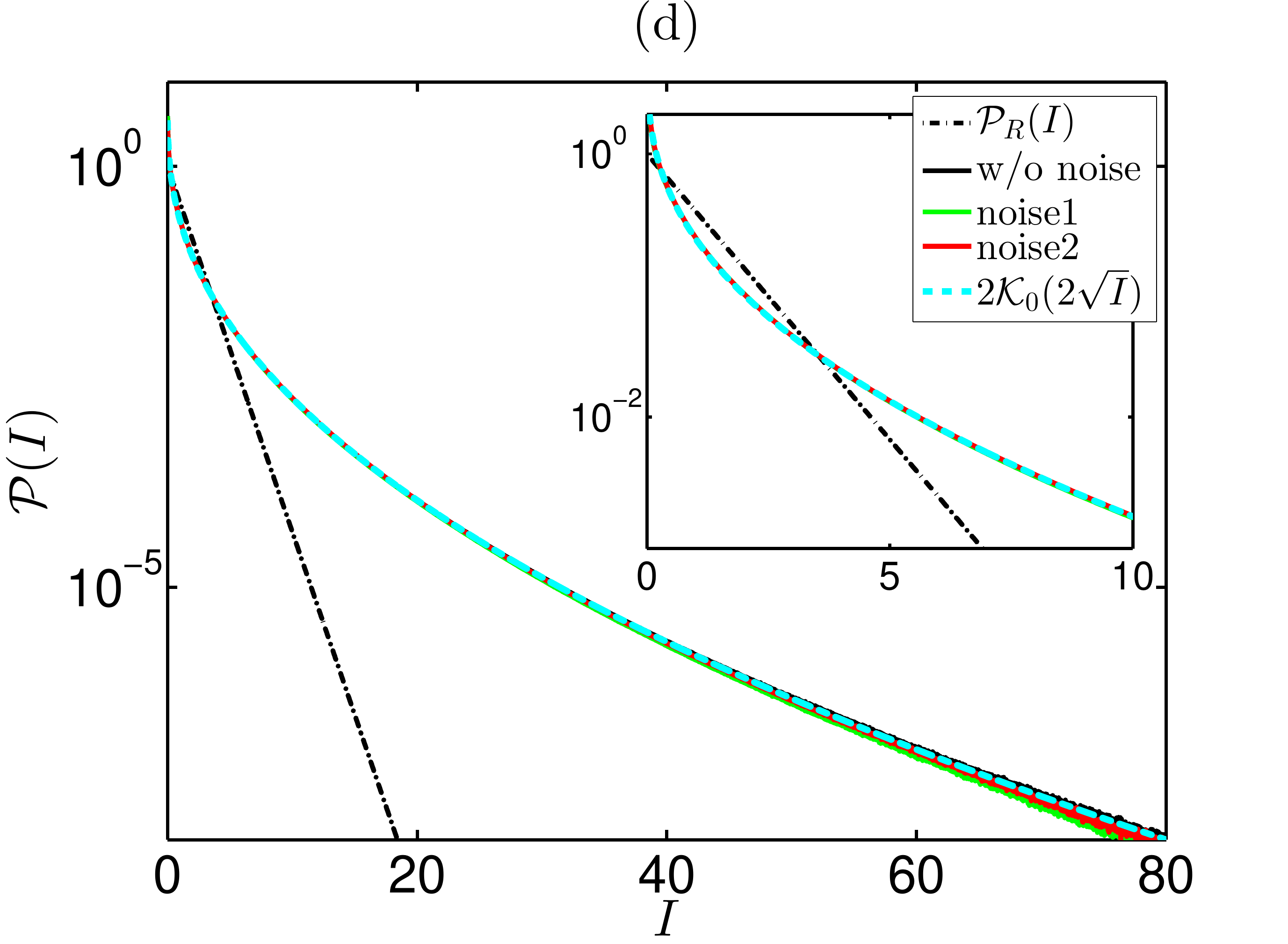}
\includegraphics[width=5.9cm]{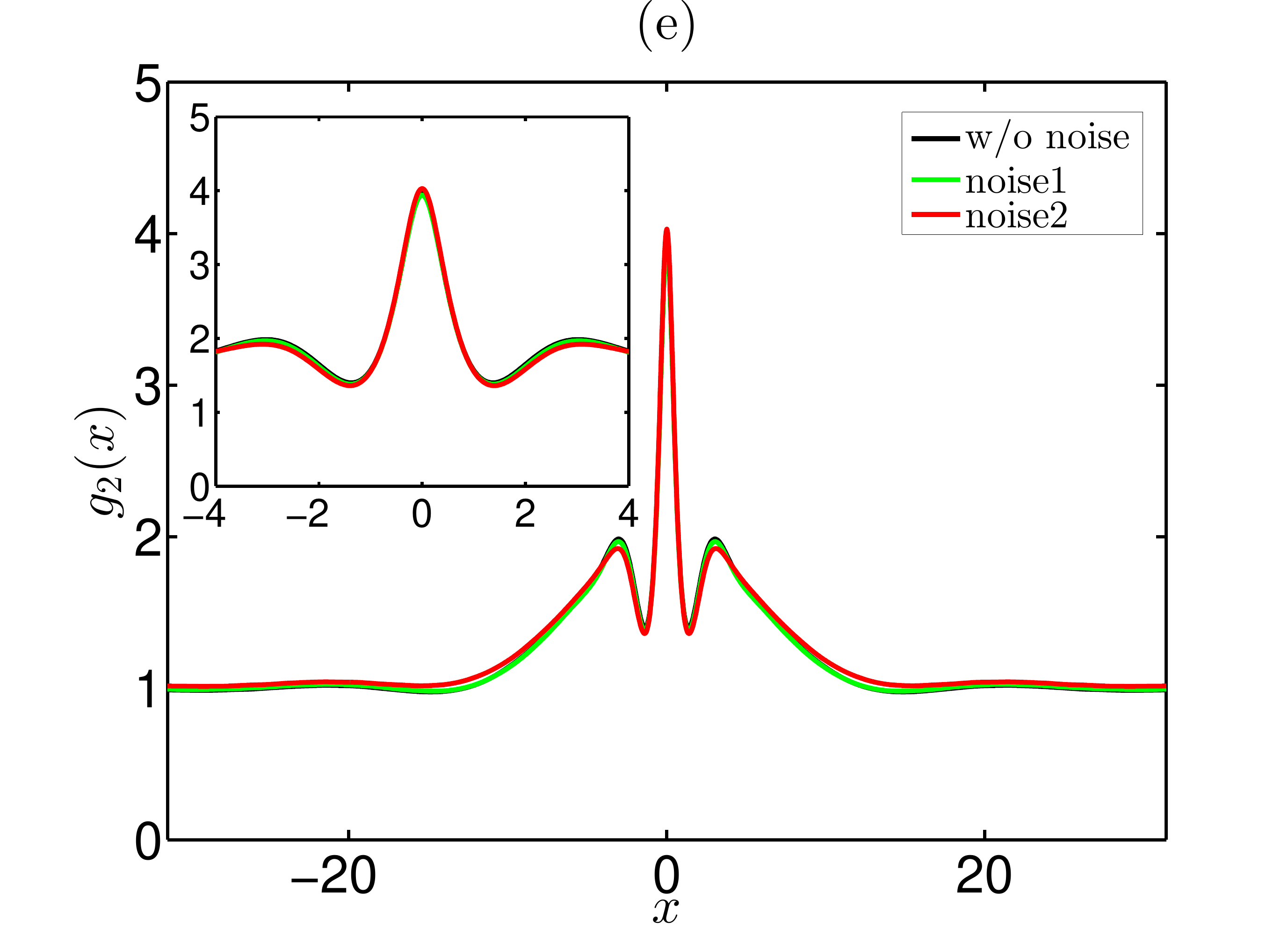}

\caption{\small {\it (Color on-line)} Influence of the additional wide-spectrum noise on the statistics. 
The black lines indicate the original experiment with super-Gaussian initial spectrum with the exponent $n=32$ and nonlinearity $\alpha_{0}=64$, the green lines -- the experiment with additional small noise~(\ref{noise-IC}) with amplitude $\chi=10^{-3}$ and super-Gaussian spectrum with $n=32$ and $\theta=4$, and the red lines -- with additional moderate noise with parameters $\chi=10^{-1}$, $n=32$ and $\theta=8$. 
(a) Evolution of the ensemble-averaged kinetic energy $\langle H_{l}\rangle$ (thin solid lines), potential energy $\langle H_{nl}\rangle$ (thick dashed lines) and the fourth-order moment $\kappa_{4}$ (thick dash-dot lines). 
(b-e) Ensemble- and time-averaged statistical characteristics in the beginning of the QSS: (b) the moments $[M^{(p)}]^{1/p}$, (c) the wave-action spectrum $S_{k}$, (d) the PDF $\mathcal{P}(I)$ and (e) the autocorrelation of intensity $g_{2}(x)$. 
The inset in the panel (b) shows the higher-order moments $M^{(p)}$ (without the $1/p$ power), the left inset in the panel (c) -- the initial wave-action spectrum at $t=0$, and the right inset in the same panel together with the insets in the panels (d-e) -- the same functions as in the main figures with smaller scales.
In the panels (d) and (b), the black dash-dot lines indicate the exponential PDF~(\ref{Rayleigh}) and the corresponding moments~(\ref{moments-Rayleigh}), while the cyan dashed lines -- the Bessel PDF~(\ref{PDF-Bessel}) and the corresponding moments~(\ref{moments-Bessel}).
}
\label{fig:fig8}
\end{figure*}

The initial conditions for both experiments represent a sum of two partially coherent waves,
\begin{equation}\label{noise-IC}
\psi_{0}(x) = \frac{\psi^{(1)}(x) + \chi\,\psi^{(2)}(x)}{\sqrt{1+\chi^{2}}},
\end{equation}
where $\psi^{(1,2)}$ are determined from Eqs.~(\ref{IC})-(\ref{IC-symmetric}), so that their average intensity is unity, $\langle|\psi^{(1,2)}|^{2}\rangle = 1$. 
In Eq.~(\ref{noise-IC}), $\psi^{(1)}$ is the (original) partially coherent wave of high nonlinearity with super-Gaussian initial spectrum with the exponent $n=32$ and nonlinearity $\alpha_{0}=64$, while $\psi^{(2)}$ is wide-spectrum wave modeling noise and $\chi$ is the average noise amplitude in the $x$-space. 
For the first experiment with small noise level, we use noise amplitude $\chi=10^{-3}$ and super-Gaussian spectrum for $\psi^{(2)}$ with $n=32$ and spectral width $\theta=4$, see Eqs.~(\ref{IC})-(\ref{IC-symmetric}). 
For the second experiment, we set these parameters at $\chi=10^{-1}$, $n=32$ and $\theta=8$. 
The left inset in Fig.~\ref{fig:fig8}(c) illustrates the initial wave-action spectrum $S_{k}$ at $t=0$ for these two additional experiments with green and red lines, in comparison with the original experiment without noise (black line). 
As in the case of the original experiment, all the statistical characteristics discussed below are averaged over ensemble of $1000$ realizations of the initial conditions, and those statistical functions that relate to the QSS -- additionally over time interval $t\in[20,40]$, which accommodates in the beginning of the QSS. 

As shown in Fig.~\ref{fig:fig8}, the results for the first experiment with additional small noise practically coincide with those for the original experiment without noise. 
The largest distinctions that we observe are very small deviations of the PDF at large intensities, together with by less than $10\%$ smaller value of the tenth-order moment $M^{(10)}$. 
We think that these discrepancies come mainly from the finiteness and randomness of the ensemble of initial conditions and disregard them. 

Stronger effects are observed for the second experiment with moderate noise level. 
In this case, the difference is seen already for the transient regime, in which the fourth-order moment starts to increase noticeably earlier and reaches smaller maximum value at slightly different time compared to the original experiment without noise, Fig.~\ref{fig:fig8}(a). 
In the QSS, the fourth-order moment and the potential energy reach the same values $\kappa_{4}\approx 4$ and $\langle H_{nl}\rangle\approx -2$ as without noise, however, the kinetic energy becomes larger, $\langle H_{l}\rangle\approx 1.2$. 
The moments practically coincide with those for the experiment without noise, Fig.~\ref{fig:fig8}(b); the measured value for the tenth-order moment $M^{10}$ turns out to be about $10\%$ smaller than without noise. 

Note that, in the presence of noise, the initial kinetic energy $\langle H_{l}\rangle|_{t=0}\approx 0.2$ is larger than that without noise approximately by the same value as in the QSS. 
The corresponding contribution to the initial kinetic energy comes from wavenumbers $|k|\simeq 8$, where the noise has intensity in the $k$-space $|\psi_{k}^{(2)}|^{2}/\Delta k\approx 6.5\times 10^{-4}$, see the left inset in Fig.~\ref{fig:fig8}(c).
Calculating the integral sum for the kinetic energy in the $k$-space in Eq.~(\ref{energy-2}), one can easily get the difference of $\simeq 0.2$ originating from this effect. 

The wave-action spectrum in the beginning of the QSS is affected by noise very slightly at small wavenumbers $|k|\lesssim 1$, and significantly at wavenumbers $|k|\simeq 8$ where the spectrum acquires ``wings'' approximately at the level $S_{k}\simeq 6\times 10^{-4}$ of the initial noise, see Fig.~\ref{fig:fig8}(c). 
These ``wings'' are the source of the larger value of the kinetic energy $\langle H_{l}\rangle$ in the QSS, compared to the experiment without noise.
The PDF and the autocorrelation of intensity turn out to be practically unaffected by noise in the second experiment, Fig.~\ref{fig:fig8}(d,e).

We conclude that a significant additional noise may noticeably change the transient regime, modify the wave-action spectrum with ``wings'' appearing at the noise level and increase the kinetic energy. 
However, even with notable noise levels, it leaves the potential energy, the moments, the PDF and the autocorrelation of intensity practically unaffected in the QSS. 
In our opinion, this opens perspectives for experiment observation of the latter statistical functions. 


\section{Conclusions}
\label{Sec:Conclusions}

In the present paper we have studied the integrable turbulence developing from partially coherent wave source of high nonlinearity. 
Motivated by the previous publications~\cite{walczak2015optical,suret2016single}, that indicated such initial conditions as ones of the most promising for the enhanced generation of rogue waves, we have examined the basic statistical characteristics of the turbulence depending on the spectrum of partially coherent wave source (its shape and nonlinearity level $\alpha_{0}$, with the latter defined as the ratio of the potential energy to the kinetic one) and also analyzed the emerging rogue waves. 

In strongly nonlinear regime, we have found that after relatively short transient the integrable turbulence enters a quasi-stationary state (QSS), while the evolution towards the long-term stationary state turns out to be very long. 
In the QSS, most of the basic statistical characteristics -- the kinetic and the potential energies, the moments, the wave-action spectrum and the PDF of intensity -- change with time very slowly, and the evolution is hidden in the wave-action spectrum at large wavenumbers, the PDF at very large intensities, the higher-order moments and the autocorrelation of intensity at the intermediate distances. 

The initial nonlinear partially coherent wave represents a collection of humps of characteristic width $\delta x\simeq \sqrt{\alpha_{0}}$, and its subsequent evolution can be subdivided by the fast motion of the smaller-scale breather-like structures developing on the background of the humps and the slow mixing and disappearance of the humps. 
For the initial nonlinearity $16\le\alpha_{0}\le 128$, all humps revealed by the spatial smoothing procedure evolve slowly. 
For larger nonlinearity $\alpha_{0}\ge 256$, the largest humps tend to significantly decrease in intensity and broaden already to the beginning of the QSS, and then split to several smaller humps, while the initially smaller humps survive and retain their shape for a long time. 
We believe that the slow mixing and disappearance of the humps is the main process that underlies the QSS and determines the slow evolution of its statistics. 
At the end of this process, we expect that the wavefield does not contain pronounced humps and the turbulence enters the asymptotic stationary state. 
However, the evolution towards the latter turns out to be very long. 
For this reason, in the present paper we focus on the detailed examination of the basic statistical characteristics in the beginning of the QSS. 

As we have found, in the QSS, the ensemble-averaged kinetic and potential energies have values $\langle H_{l}\rangle\approx 1$ and $\langle H_{nl}\rangle\approx -2$, so that the ratio of the potential energy to the kinetic one is close to two. 
The value of the fourth-order moment $\kappa_{4}\approx 4$ indicates to a strongly non-exponential PDF and enhanced generation of rogue waves. 
The three statistical characteristics $\langle H_{l}\rangle$, $\langle H_{nl}\rangle$ and $\kappa_{4}$ practically do not change during the QSS, and we think that in the asymptotic stationary state they have the same values. 

In the beginning of the QSS, the wave-action spectrum is symmetric, decays slightly slower than exponential at large wavenumbers $|k|\gtrsim 4$, and has a triangular shape at small wavenumbers, $|k|\lesssim 1$. 
The spectrum does not depend on its initial shape, even when the latter is rather generic and nonsymmetric, but changes significantly with the level of the initial nonlinearity $\alpha_{0}$. 
In particular, the triangular profile at small wavenumbers $|k|\lesssim 1$ sharpens with increasing $\alpha_{0}$, resembling the asymptotic spectrum of the noise-induced MI~\cite{agafontsev2015integrable} for large $\alpha_{0}$. 

The PDF of relative wave intensity is slightly smaller than the exponential distribution~(\ref{Rayleigh}) at moderate intensities $I\simeq 2$, and exceeds it by orders of magnitude at large intensities $I\gtrsim 5$. 
The PDF does not depend on the shape of the initial spectrum too, and we have found an interval of the initial nonlinearity $16\lesssim\alpha_{0}\lesssim 128$, where the PDF practically does not change with $\alpha_{0}$ as well. 
The universal shape of the PDF in this interval is very well approximated by a Bessel function~(\ref{PDF-Bessel}). 
The latter is derived under the assumptions that the humps change with time very slowly, and, for every small element of a hump, the intensity of the smaller-scale breather-like structures evolving on its background is exponentially distributed. 

When the initial nonlinearity is not large enough $\alpha_{0}\lesssim 4$, the humps have characteristic width close to that of the breather-like structures, and we observe the PDF noticeably smaller than the Bessel PDF~(\ref{PDF-Bessel}) at large intensities. 
For $\alpha_{0}\gtrsim 256$, the largest humps significantly decrease and broaden already to the beginning of the QSS, and the PDF at large intensities also deviates from the Bessel PDF to slightly smaller values. 
While for $\alpha_{0}=256$ these deviations start to get visible from $I\gtrsim 50$ and remain relatively small, we expect them to increase with $\alpha_{0}$, since the described tendency for the largest humps to decrease and broaden spreads to smaller and smaller humps with increasing initial nonlinearity. 

The moments of amplitude $M^{(p)}$ have universal dependency on their order $p$ for $16\lesssim\alpha_{0}\lesssim 128$, and this dependency is very well approximated by the moments~(\ref{moments-Bessel}) derived from the Bessel PDF~(\ref{PDF-Bessel}). 
Rather remarkably, the Bessel moments equal to the squares of the moments~(\ref{moments-Rayleigh}) following from the exponential PDF~(\ref{Rayleigh}). 
For $\alpha_{0}\lesssim 4$ and $\alpha_{0}\gtrsim 256$, we observe deviations of the higher-order moments from Eq.~(\ref{moments-Bessel}) to smaller values. 

The autocorrelation of intensity has universal bell-shaped central part with full width at half maximum $\Delta_{FWHM}\simeq 1.4$, and also non-universal part at intermediate distances between the universal central peak and the large distances where the autocorrelation reaches unity. 
The non-universal part depends on both the shape and the nonlinearity level of the initial spectrum, and occupies interval of distances that expands roughly as $\sqrt{\alpha_{0}}$ with increasing $\alpha_{0}$. 
The maximum of the autocorrelation function equals to the fourth-order moment, $\max g_{2}(x)=g_{2}(0)=\kappa_{4}\approx 4$.

The wave-action spectrum, the PDF, the moments, and the autocorrelation of intensity change with continued evolution in time. 
In particular, at sufficiently long time, the most noticeable changes are observed for the wave-action spectrum at small wavenumbers, where its triangular shape becomes less sharp, and at large wavenumbers, where the spectrum decreases. 
This process is accompanied by the PDF decreasing visibly at large intensities and by the decreasing higher-order moments, and also by the non-universal part of the autocorrelation function shrinking to the width of unity order. 

A feasible way to create strongly nonlinear partially coherent waves is to narrow the initial spectrum; however, in real physical systems an additional wide-spectrum noise may be present. 
Our comparison of the experiments with and without additional noise has shown that small noise has negligible effect on the statistics. 
Larger noise may noticeably change the transient regime and modify the wave-action spectrum in the QSS with ``wings'' appearing at the noise level; these wings lead to increase in the kinetic energy.
However, even with notable noise levels, many of the basic statistical functions in the QSS -- the potential energy, the PDF, the moments and the autocorrelation of intensity -- are practically unaffected by the noise presence.

The value of the fourth-order moment $\kappa_{4}\approx 4$, as well as the PDF exceeding the exponential distribution by orders of magnitude at large intensities, indicate enhanced appearance of large waves. 
In particular, we have detected rogue waves with amplitude by more than $14$ times larger than the average one. 
So large an excess comes from the following behavior of the rogue waves: the initial conditions contain humps exceeding the average amplitude by several times, and the evolution of the breather-like structures on the background of the humps leads to spikes exceeding the amplitude of the humps by several times more. 
We have systematically examined the largest rogue waves detected in the beginning of the QSS for each of our numerical experiments with initial nonlinearity $\alpha_{0}\ge 4$ and different profiles of the initial spectrum, and found that all of these rogue waves are very well approximated -- both in space and in time -- by the rational breather solutions of either the first (the Peregrine breather), or the second orders. 
The space-time representation of the observed rogue waves looks like a collision of several pulses with width of unity order. 

The integrable turbulence developing from partially coherent wave source of high nonlinearity evolves towards the asymptotic stationary state for a very long time, and in the present paper we were not able to examine this final state. 
For the latter, however, we think that an alternative strategy based on the inverse scattering transform might be useful. 
The main role in the statistics of an initially highly nonlinear wavefield should be played by its solitonic content, see e.g.~\cite{novikov1984theory} and also the latest study~\cite{gelash2019bound} of the noise-induced MI based on the soliton gas approach. 
Then, identification of the scattering data for the solitonic content, see~\cite{mullyadzhanov2019direct,gelash2019direct} and the references wherein, and the subsequent reconstruction of the soliton gas~\cite{gelash2018strongly} with the same distribution of the IST eigenvalues, but with random phases for the so-called norming constants, may result in a very good approximation of the asymptotic stationary state. 
We are going to explore this approach in the near future. 

Our main motivation for this study was the fundamental investigation of the integrable turbulence in a setting that allows frequent appearance of very large rogue waves. 
However, our results may be useful for practical applications as well, in particular, either as a construction method for a wavefield with frequent spikes exceeding the average power by several hundreds times, or, alternatively, as a possible source of large spikes that needs to be eliminated.


\begin{center}
\textbf{Acknowledgements}
\end{center}

The authors thank A. Gelash, A. Tikan and F. Copie for fruitful discussions. 
Simulations were performed at the Novosibirsk Supercomputer Center (NSU). 
This work has been partially supported by the Agence Nationale de la Recherche through the LABEX CEMPI Project (Project No. ANR-11-LABX-0007) and the I-SITE ULNE (ANR-16-IDEX-004) and by the Ministry of Higher Education and Research, Hauts-De-France Regional Council and European Regional Development Fund (ERDF) through the Contrat de Projets Etat-Region (CPER Photonics for Society Grant No. P4S). 
The work of D.S.A. (simulations) was supported by the state assignment of IO RAS, Grant No. 0149-2019-0002.


\appendix
\section{Non-symmetric initial spectrum}
\label{Sec:AppA}

We construct generic (non-symmetric) Fourier spectrum $A_{k}^{(0)}$ in Eq.~(\ref{IC}) in the following way. 
At the first step, we generate an ``intermediate'' spectrum
\begin{eqnarray}\label{IC-nonsymmetric-spectrum-1}
&&\tilde{A}_{k}^{(0)} = C_{0}\exp\bigg[f_{k} - (1-h_{k})\bigg(\frac{|k|}{\theta_{l}}\bigg)^{n_{l}} -\nonumber\\ 
&&-h_{k}\bigg(\frac{|k|}{\theta_{r}}\bigg)^{n_{r}}\bigg],
\end{eqnarray}
where $C_{0}$ is a constant determined from the normalization condition $\overline{|\psi_{0}(x)|^{2}}=1$ (equivalent to $\sum_{k}[\tilde{A}_{k}^{(0)}]^{2}=1$), $f_{k}\sim 1$ is generic real function and $h_{k}$ is the Heaviside step function. 
The ``left'' and ``right'' decay exponents $n_{l}$ and $n_{r}$ together with the ``left'' and ``right'' coefficients $\theta_{l}$ and $\theta_{r}$ enable the spectrum $\tilde{A}_{k}^{(0)}$ to decay differently in the limits $k\to -\infty$ and $k\to +\infty$. 
The function $f_{k}$ is composed as a superposition of linear waves with Gaussian spectrum and arbitrary phases (in the same way as we define the initial conditions~(\ref{IC})-(\ref{IC-symmetric})). 

\begin{figure}[t]\centering
\includegraphics[width=7.5cm]{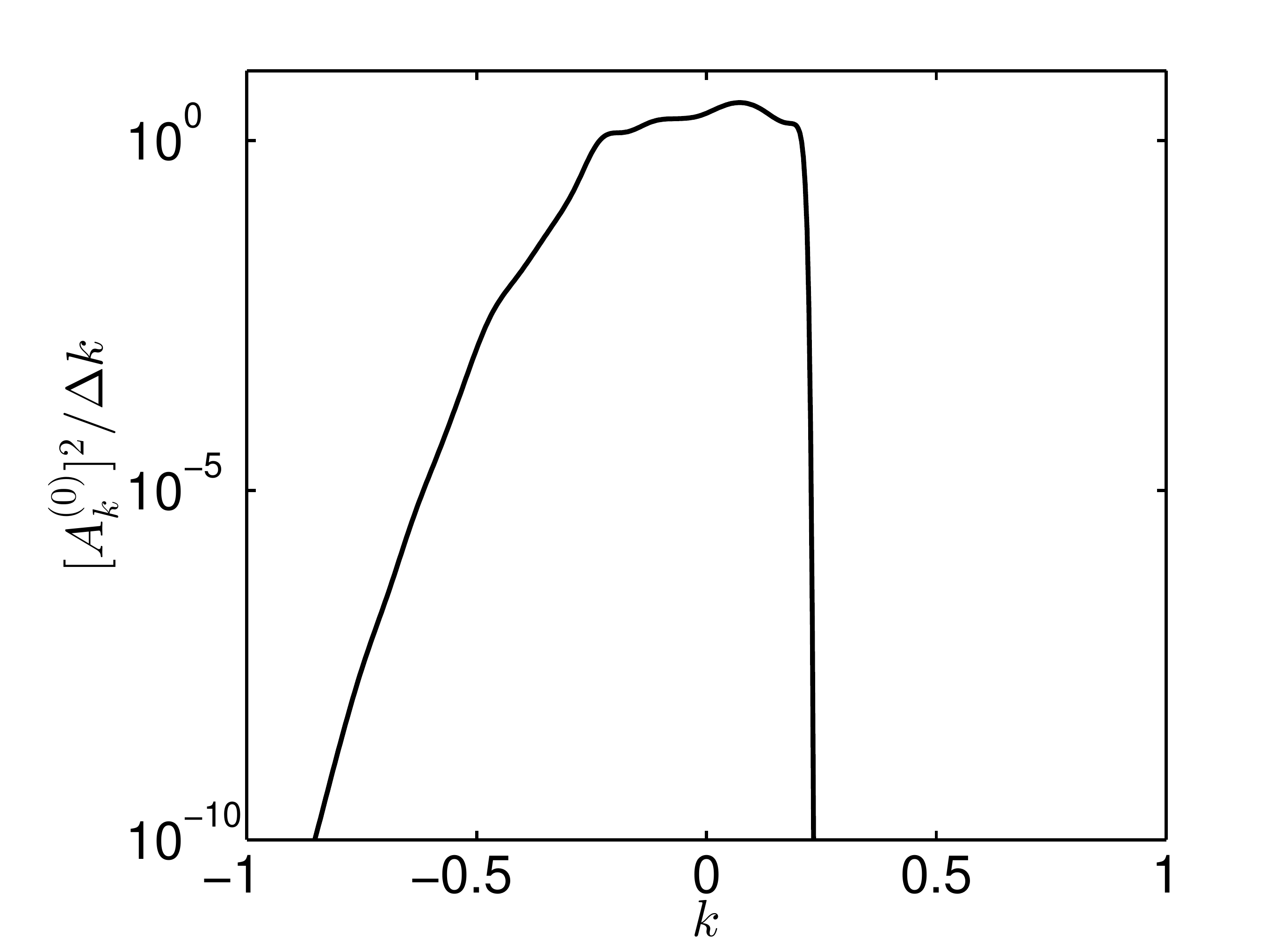}

\caption{\small 
Non-symmetric initial wave-action spectrum $[A_{k}^{(0)}]^{2}/\Delta k$ for one of the numerical experiments. The spectrum has ``left'' and ``right'' decay exponents $n_{l}=2$ and $n_{r}=32$, and equal coefficients $\theta_{l}=\theta_{r}=0.24$, see Eqs.~(\ref{IC-nonsymmetric-spectrum-1})-(\ref{IC-nonsymmetric-spectrum-2}), and is characterized by the ensemble-average potential-to-kinetic energy ratio $\alpha_{0}\approx 65.2$. 
}
\label{fig:fig9}
\end{figure}

The ``intermediate'' spectrum has non-zero momentum~(\ref{momentum}), $\sum_{k}k[\tilde{A}_{k}^{(0)}]^{2}\neq 0$, and, in order to eliminate it, at the second step we construct the spectrum $A_{k}^{(0)}$ (which is used in Eq.~(\ref{IC}) for generation of the initial conditions) by shifting the ``intermediate'' spectrum by a constant wavenumber $k_{0}$, 
\begin{equation}\label{IC-nonsymmetric-spectrum-2}
A_{k}^{(0)} = \tilde{A}_{k-k_{0}}^{(0)}, \quad k_{0} = -\sum_{k}k [\tilde{A}_{k}^{(0)}]^{2}.
\end{equation}
For the function $f_{k}$, this shift is performed with the (forward) Fourier transform to the $x$-space, multiplication by $e^{-ik_{0}x}$ and then the (backward) Fourier transform to the $k$-space. 
One of the examples of non-symmetric spectrum $A_{k}^{(0)}$ used for the numerical experiments of this paper is shown in Fig.~\ref{fig:fig9}. 
It has ``left'' and ``right'' decay exponents $n_{l}=2$ and $n_{r}=32$, and equal coefficients $\theta_{l}=\theta_{r}=0.24$, and is characterized by the ensemble-average potential-to-kinetic energy ratio $\alpha_{0}\approx 65.2$. 
For our simulation parameters, in the region $[-\theta_{l},\theta_{r}]$ this spectrum is resolved with $(\theta_{l}+\theta_{r})/\Delta k\sim 100$ harmonics, where $\Delta k = 2\pi/L = 1/256$, $L=512\pi$, is the distance between neighbor harmonics.

\bibliographystyle{apsrev4-1}
\bibliography{refs}

\end{document}